\documentclass[10pt,conference]{IEEEtran}
\IEEEoverridecommandlockouts
\usepackage{cite}
\usepackage{amsmath,amssymb,amsfonts}
\usepackage{algorithmic}
\usepackage{graphicx}
\usepackage{textcomp}
\usepackage{tcolorbox}
\usepackage{todonotes}
\usepackage{url}
\usepackage{listings}
\usepackage{booktabs}
\usepackage{xspace}
\usepackage{hyperref}

\lstdefinestyle{mystyle}{
    basicstyle=\ttfamily\footnotesize,
    language=Java,
    captionpos=b,                    
    showspaces=false,                
    showstringspaces=false,
    tabsize=2
}
\newcommand{\urlRepo}[0]{\url{https://osf.io/yh6v2/?view_only=a069f5a56f3640abaf235eed1320b46f}\xspace} 

\usepackage[inline]{enumitem}
\setlist[enumerate,1]{label=\textit{\alph*)}}
\definecolor{review}{RGB}{0,0,0}

\usepackage{xcolor}
\def\BibTeX{{\rm B\kern-.05em{\sc i\kern-.025em b}\kern-.08em
    T\kern-.1667em\lower.7ex\hbox{E}\kern-.125emX}}

\usepackage{booktabs}

\begin{document}

\title{On the Possibility of Breaking Copyleft Licenses When Reusing Code Generated by ChatGPT
%\title{Investigating Possible IP Rights Infringement in AI-Generated Code: A Case Study on ChatGPT\\
%{\footnotesize \textsuperscript{*}Note: Sub-titles are not captured for https://ieeexplore.ieee.org  and
}

\author{\IEEEauthorblockN{Gaia Colombo}
\IEEEauthorblockA{
\textit{University of Milano-Bicocca}\\
Milano, Italy \\
g.colombo147@campus.unimib.it}
\and
\IEEEauthorblockN{Leonardo Mariani}
\IEEEauthorblockA{
\textit{University of Milano-Bicocca}\\
Milano, Italy \\
leonardo.mariani@unimib.it}
 \and
%\linebreakand
%\hspace{-1.0cm} % Adjust spacing between authors if needed
\IEEEauthorblockN{Daniela Micucci}
\IEEEauthorblockA{
\textit{University of Milano-Bicocca}\\
Milano, Italy \\
daniela.micucci@unimib.it}
\and
\IEEEauthorblockN{Oliviero Riganelli}
\IEEEauthorblockA{
\textit{University of Milano-Bicocca}\\
Milano, Italy \\
oliviero.riganelli@unimib.it}

}

%\author{\IEEEauthorblockN{Anonymous Authors}}
%\and
%\and
%\and
%\and
%\and

\maketitle

\begin{abstract}
AI assistants can help developers by recommending code to be included in their implementations (e.g., suggesting the implementation of a method from its signature). Although useful, these recommendations may mirror copyleft code available in public repositories, exposing developers to the risk of reusing code that they are allowed to reuse only under certain constraints (e.g., a specific license for the derivative software). 

This paper presents a large-scale study about the frequency and magnitude of this phenomenon in ChatGPT. In particular, we generate more than 70,000 method implementations using a range of configurations and prompts, revealing that a larger context increases the likelihood of reproducing copyleft code, but higher temperature settings can mitigate this issue.
 %Our findings reveal that a larger matching context increases the likelihood of GPT-4 reproducing copyleft code, though higher temperature settings can help mitigate this issue.

\end{abstract}


\section{Introduction}

Several AI assistants, such as ChatGPT\footnote{https://openai.com/chatgpt/}, GitHub Copilot\footnote{https://github.com/features/copilot}, and Google Gemini\footnote{https://gemini.google.com/}, are available to help developers complete their development tasks. Multiple studies reveal that, although far from perfect, these tools may produce useful results with the potential of accelerating code development~\cite{Hou:LLMforSE:SLR:TOSEM:2024,Fan:LLMChallenges:ICSEFOSE:2023,Mastropaolo:RobustnessGenCode:ICSE:2023,Corso:EmpiricalAssessment:ICPC:2024,Fagadau:EmpiricalStudy:ICPC:2024}. 

AI assistants exploit LLMs trained on a huge corpus of data, including code, to provide recommendations. This raises concerns in terms of the \emph{ownership and rights to reuse} the code generated by these tools when the recommended code matches the code in the training set. This issue is particularly severe if we consider that training code could be protected by restrictive licenses that forbid or limit reuse\footnote{For instance, open source code protected by copyleft licenses can be reused only under specific constraints on how the derivative code is licensed.}. That is, a developer may inadvertently break the license terms associated with the reused code by simply accepting a recommendation.

A key question about the recommendations produced by AI assistants is thus: \emph{``Is the recommended code original, or is it a copy of existing code protected by a restrictive license?"} Answering this question can be extremely important to prevent potentially unethical behavior and possible legal issues. 

Previous studies based on the now-obsolete GPT-2 models suggest that LLMs can memorize and reproduce code present in their training data. In particular, Al-Kaswan et al.~\cite{AlKaswan:TracesMemorisation:ICSE:2024}  reported that LLMs may generate completions for code prefixes encountered during training by appending the corresponding code suffix from the training data.
%OLD
%Previous studies based on, now obsolete, GPT-2 models show that LLMs may memorize and reproduce the code that is part of the training data. In particular, Al-Kaswan et al.~\cite{AlKaswan:TracesMemorisation:ICSE:2024} reported that LLMs may suggest to complete a code-prefix that occurred in the training data, with the suffix-code present in the training data. 
Similarly, Yang et al.~\cite{Yang:UnveilingMemorization:ICSE:2024} found that training code can be observed in the output for a range of diverse prompts. 

A more recent study~\cite{Yu:codeipprompt:ICML:2023} investigated the concern of code memorization in GPT-4 models, considering code protected by restrictive licenses. The investigation reported a high probability (more than 50\%) for models to return code protected by copyleft licenses if queried for the generation of a function's implementation whose signature is present in the training set. 

%Such a high percentage of potential violations is, also, due to the configuration of the plagiarism detection tools used in the study. In fact, the code returned by AI-assistants is considered as a potential evidence of plagiarism if its similarity is above 0.5 for at least one of the two plagiarism detection tools used: JPlag and Devos. 

%This setup is in contrast with previous studies~\cite{AlKaswan:TracesMemorisation:ICSE:2024,Yang:UnveilingMemorization:ICSE:2024} that considered nearly identical code as an evidence of plagiarism (e.g., Type I clones), and reported a definitely lower probability (\textbf{LEO quanto?}) for AI-assistants to generate code that plagiarizes code in the training data. 

This study aims to \emph{broaden existing evidence and derive further insights} about the risks related to the reuse of the code originated by GPT models. In particular, we investigate how GPT-4 performs in the generation of Java methods, studying the \emph{similarity} between the recommended code and the corresponding open source code available under restrictive licenses using the JPlag plagiarism detection tool\footnote{https://helmholtz.software/software/jplag}. Further, our investigation considers multiple dimensions not studied in the context of GPT-4 models, so far. 

First, we study if and to what extent the \emph{context} may influence the generation of code protected by restrictive licenses. In particular, %we consider the case of developers who have already accepted some code recommendations that include licensed code, and we study if this may increase the likelihood of receiving further recommendations of code still protected by restrictive licenses. 
we investigate cases in which developers, having already accepted recommendations containing licensed code, may face an increased likelihood of receiving additional recommendations of code protected by restrictive licenses. We specifically consider the case of classes that already contain code that mirrors open-source code protected by copyleft licenses, and study its impact on recommendations.
We investigate two paradigmatic cases: one considering the presence of \emph{all the methods in the class but the one that has to be generated} matching an open-source implementation protected by a copyleft license, and another one only considering the \emph{presence of access methods} matching an open-source implementation protected by a copyleft license. The first case captures the situation where the context is the largest possible, encompassing every other method within the class. Instead, the second case investigates to what extent the presence of simple and commonly implemented methods with minimal semantic content (i.e., the access methods) may influence the generation of licensed code.

Second, we consider the impact of the \emph{temperature} on the likelihood of generating code protected by restrictive licenses. The temperature is a parameter that influences the creativity of the model, and thus it may have a role in the likelihood that the model replicates licensed code observed during training.

%Last, we investigate if GPT-4 shows any level of \emph{awareness} in the generation of code distributed under restrictive licenses, by studying if explicitly asking for original code may decrease the change code that cannot be freely reused is recommended. 

Finally, we examine if GPT-4 demonstrates any level of \emph{awareness} regarding the generation of code subject to restrictive licenses. Specifically, we investigate whether explicitly requesting code that is not a copy of any known implementation may influence the likelihood of receiving recommendations for code that cannot be freely reused.

We performed this investigation on a body of 7,347 methods, assessing more than 70,000 method implementations. Results show a low probability of receiving recommendations of copyleft code for individual requests, but an increasingly large matching context may increase the likelihood that GPT-4 returns copies of code protected by copyleft licenses. 
%That is, the 
This suggests that the early acceptance of copyleft code may lead developers to incidentally accumulate more copyleft code in their implementations. High-temperature values may help mitigate this problem. Finally, GPT-4 has not demonstrated any awareness of this phenomenon.

In a nutshell, this paper contributes as follows:
\begin{enumerate*}[leftmargin=*,label=\textit{\alph*)}]
\item Reporting a large-scale study on the generation of licensed code with GPT-4,

\item Presenting an analysis of how the context and the temperature may influence the results,

\item Investigating if any trace of awareness is present in GPT-4 models concerning the generation of code under restrictive licenses,

\item Producing recommendations for developers who use GPT-4 to generate code,

\item Publicly releasing the experimental material\footnote{\urlRepo} for replication and extension of the study.

\end{enumerate*}

The paper is organized as follows. Section~\ref{sec:RW} discusses related work. Section~\ref{sec:Met} introduces the five research questions investigated in this paper and presents the methodology used to answer them. Section~\ref{sec:Res} reports the empirical results collected to answer the five research questions. Finally, Section~\ref{sec:con} provides final remarks and suggests directions for future research.

\section{Related Work}
\label{sec:RW}
\subsection{Code Generation with LLM}
The advent of LLMs has significantly transformed the landscape of software engineering, particularly in code generation \cite{Fan:LlmSeSurveyProblems:ICSE-FoSE:2023}, reflecting the growing demand for tools that enhance software development efficiency \cite{Hou:SurveyLlmSe:TOSEM:2024}.

%Code generation, the process of converting natural language descriptions into executable source code, has gained significant attention due to its potential to simplify programming tasks and make coding more accessible to a broader audience. 
Recent advancements have demonstrated impressive capabilities in translating natural language prompts into functional code snippets, and these models have been used to address various coding tasks \cite{Li:LLMLoggingStatement:TSE:2024,Schafer:LLMUnitTesting:TSE:2023,Xia:LLMAutomatedProgramRepair:ICSE:2023}.
%
%For example, one significant area of exploration is using LLMs for automated unit test generation. \cite{Schafer:LLMUnitTesting:TSE:2023} highlights the capabilities of LLMs in generating test cases. However, it also points out that the executability of these tests can be inconsistent, indicating a need for further refinement in the generation process to ensure that the produced tests are functional and reliable. In addition to testing, LLMs have been effectively utilized for generating logging statements in code. 
%
Indeed, they have been extensively used to generate functional code~\cite{Yetistiren:AssessingCopilot:PROMISE:2022,Corso:EmpiricalAssessment:ICPC:2024,Vaithilingam:ExpExpCoGen:CHI:2022,Mastropaolo:RobustnessGenCode:ICSE:2023,liu2024your,Fagadau:EmpiricalStudy:ICPC:2024,Donato:LLMConfig:ICPC:2025}, with studies showing that while the generated code is often syntactically correct, it may lack semantic accuracy, requiring human intervention for validation and refinement. %Current studies reveal that, while the generated code is often syntactically correct, it may lack semantic accuracy and require human intervention for validation and refinement. 

%OLD:%Additionally, while generated code can often meet the requirements of simple tasks, it may struggle with more complex scenarios.

In addition to code generation, LLMs have also been used for other tasks. For example, the work by Li et al~\cite{Li:LLMLoggingStatement:TSE:2024} shows how LLMs can assist developers by automatically inserting logging code, which is crucial for debugging and maintaining software applications. 
Xia et al.~\cite{Xia:LLMAutomatedProgramRepair:ICSE:2023} discuss how LLMs can be leveraged to suggest fixes, thereby streamlining the debugging process and reducing the workload on developers.
LLMs have been also used to automatically generate unit test cases~\cite{Schafer:LLMUnitTesting:TSE:2023}.

%The quality of the code generated by LLMs has also been critically assessed in several studies \cite{Yetistiren:AssessingCopilot:PROMISE:2022,Corso:EmpiricalAssessment:ICPC:2024}. These studies reveal that while the generated code is often syntactically correct, it may lack semantic accuracy and require human intervention for validation and refinement. Additionally, while generated code can often meet the requirements of simple tasks, it may struggle with more complex scenarios.

The body of work surrounding LLMs in code generation is rapidly expanding, with ongoing research addressing various applications, challenges, and future directions. As these models evolve, they hold the potential to significantly transform software engineering practices \cite{Fan:LlmSeSurveyProblems:ICSE-FoSE:2023,Hou:SurveyLlmSe:TOSEM:2024}.

This study complements these findings with evidence about the potential risks
regarding the possible inclusion of copyleft licensed code in the recommended code.

\subsection{Plagiarism in AI-generated Code}
Recent studies have increasingly focused on the implications of content generation by LLMs and the potential issues about originality. Some studies have specifically addressed these issues in the context of code generated by LLMs \cite{AlKaswan:TracesMemorisation:ICSE:2024,Yang:UnveilingMemorization:ICSE:2024,Ciniselli:CloningCode:ICPC:2022,Yu:codeipprompt:ICML:2023}. 

Al-Kaswan et al. \cite{AlKaswan:TracesMemorisation:ICSE:2024} investigate the extent to which large code models memorize their training data, revealing that a significant number of outputs can reproduce memorized code snippets. The findings indicate a strong correlation between the frequency of code snippets in the training data and their occurrence in model outputs, with larger models and longer outputs exhibiting greater memorization. %Additionally, the study highlights potential privacy and security risks associated with this memorization, emphasizing the need for strategies to mitigate these issues.
Similarly,  Yang et al. \cite{Yang:UnveilingMemorization:ICSE:2024} investigate the extent to which large code models memorize their training data, revealing that a significant number of outputs can reproduce memorized code snippets. 

Ciniselli et al. \cite{Ciniselli:CloningCode:ICPC:2022} investigate the degree to which Deep Learning (DL) code recommenders, specifically a T5 model trained on over 2 million Java methods, generate code snippets that are clones of their training data. %Using the Simian clone detector, 
The researchers found that approximately 10\% of the generated predictions represented Type-1 clones, while around 80\% were Type-2 clones. However, the likelihood of generating clones decreased significantly when the model produced more complex predictions. %, with virtually no clones identified in outputs consisting of at least four lines of code. 
These findings highlight the potential for DL-based code recommenders to suggest original code while also raising important considerations regarding licensing and the originality of generated outputs.

These studies focus specifically on code cloning and memorization, raising the critical concern of plagiarism in AI-generated code, which is addressed by our study.
%They highlight the complexities surrounding plagiarism in this context and emphasize the need for customized methodologies that address the unique challenges posed by LLMs. Our work focuses on the legal implications of code generation by ChatGPT, particularly regarding the risks of plagiarism from restrictive licenses and the effectiveness of explicit requests for original content generation.

Related to our work, {\small \sc CODEIPPROMPT}~\cite{Yu:codeipprompt:ICML:2023} is a framework designed to evaluate the potential intellectual property (IP) infringement in code generated by LLMs. The analysis conducted by the authors shows that models may frequently generate code that closely resembles copyrighted material. The findings highlight significant risks of IP violations due to the presence of copyrighted code in training datasets, underscoring the need for improved data management and mitigation strategies to address these concerns effectively. %Overall, the work emphasizes the importance of reconsidering training data and developing more intelligent models to enhance IP protection in code generation. 
Our work extends this study %on plagiarism detection in AI-generated code 
by offering additional insights on plagiarism concerns that may affect AI-generated code. 
We explored various aspects that influence code generation, such as the context and the model's creativity, understanding how they affect the likelihood of obtaining code that is similar to copyleft code. 

%By analyzing these factors, our study aims to provide deeper insights into the mechanisms behind code generation, ultimately contributing to the development of more effective strategies for mitigating plagiarism risks in AI-generated code.

\section{Methodology}
%This section outlines the research questions that motivate our study and the methodology used to address these questions.
\label{sec:Met}
The primary aim of this research is to investigate the potential for ChatGPT to recommend code protected by restrictive licenses. In particular, we formulated the following research questions (RQs):

    \noindent  \textbf{RQ1: Can ChatGPT return code protected by copyleft licenses?} This question aims to determine whether ChatGPT can generate code subject to limitations imposed by copyleft licenses. The focus is on identifying actual instances of code plagiarism that may (accidentally) lead to improper reuse. Understanding this risk is important for developers who want to use AI-generated code responsibly and in compliance with licensing rules.

    %\noindent  \textbf{RQ2: Does the context provided in the request influence the likelihood of returning code protected by copyleft licenses?} This question investigates if the presence of code already processed at training time in the prompt may affect the likelihood of obtaining code protected by copyleft licenses. In particular, we consider the case of a class that has been already partially implemented consistently with publicly available copyleft code (e.g., by formerly accepting code recommendations that match with the code in an existing class), and we want to establish if this affects the likelihood of receiving recommendations of code in that same class. \textcolor{review}{The goal is to establish understand if providing larger code snippets increases the chances of generating plagiarized content.}

    \noindent  \textbf{RQ2: Does the context provided in the request influence the likelihood of returning code protected by copyleft licenses?} This question examines whether the presence of copyleft code already included in a partially implemented class increases the probability that ChatGPT will recommend additional copyleft code for that class. The rationale for this investigation stems from the concern that once a developer accepts an initial recommendation that includes copyleft code, subsequent recommendations might be affected. Understanding this dynamic is crucial, as it highlights the potential for a cascading effect where the initial inclusion of copyleft code leads to more copyleft code recommended, thereby increasing the risk of unintentional copyleft violations.

    \noindent  \textbf{RQ3:  Does providing a context with only access methods affect the likelihood of obtaining code protected by copyleft licenses?} This question investigates a specific case of context, that is, whether the presence of simple non-characteristic methods, access methods in particular, influences the likelihood  ChatGPT recommends code from classes protected by a copyleft license. Unlike RQ2, which focuses on the recommendations generated for classes with a large context already matching existing copyleft code, RQ3 focuses on the impact of a matching context that carries very limited class semantics. We specifically consider the case of access methods, which serve as getters and setters, since they convey little information about the class's overall functionality. It is important to determine if such a constrained, and apparently harmless, context can inadvertently lead to further recommendations of copyleft code.

    \noindent  \textbf{RQ4: Does adjusting the temperature parameter alter the likelihood of obtaining code protected by copyleft licenses?} This question aims to explore whether modifying the temperature, which is a parameter that controls the level of creativity of the model, can change the likelihood that the generated code matches code protected by restrictive licenses.  By exploring this relationship, we aim to determine whether a higher or lower temperature setting can effectively reduce the risk of producing code that may inadvertently infringe copyleft licenses. 

    \noindent  \textbf{RQ5: Is ChatGPT aware of using protected code and can it avoid plagiarism when explicitly requested?} 
    This question investigates whether ChatGPT can refrain from providing copied code and instead generate original content, not copied from any source, when explicitly asked to do so. The analysis aims to determine if an explicit request is effective in reducing the risk of obtaining plagiarized content.

The rest of this section describes the methodology that we used to answer these five RQs. 

\subsection{Dataset Construction}
To answer our RQs, we created a dataset of copyleft Java methods with JavaDoc comments extracted from publicly available repositories on GitHub~\cite{GitHub2024}. This dataset serves as a foundation to query ChatGPT for the generation of method implementations and checking if the recommendation matches with existing copyleft code. To select these methods, we referred to the criteria below.

\textbf{Repository Selection.} In our repository selection process, we established several key criteria to ensure that the repositories we analyzed were relevant, high-quality, and suitable for our investigation of code recommendations that may break copyleft licenses.

Firstly, we considered the licenses under which the repositories were released. We specifically selected repositories protected by any of the following open-source licenses: GPL, AGPL-3.0, CC-BY-SA-4.0, ECL-2.0, and EUPL-1.1. These licenses allow for the use and modification of the code while adding restrictions on how the derivative code can be made available and distributed (e.g., a specific license must be used). This criterion was crucial to ensure that the code we considered could not be reused without restrictions.
%analyzed could be legally reused and modified, aligning with the ethical considerations of our research. 

Second, to make sure ChatGPT could have processed during training the methods we ask it to generate, we select only methods created \emph{before} the last training date of the model used in this work.
According to official documentation, %the \texttt{gpt-3.5-turbo} model was trained until September 2021, while 
the \texttt{gpt-4-turbo} model was trained in 2023, so we opted for selecting repositories that were already existing by the end of 2020. %, taking into account the necessary time for data collection and model training.

Third, to select popular projects appreciated by the community, we included only those repositories that had at least 500 stars on GitHub. This threshold indicated a substantial level of interest and validation from the community. % suggesting that the code was well-regarded and likely of higher quality. 
By concentrating on popular repositories, we aimed to enhance the relevance and reliability of our analysis.

Fourth, since we target Java in our analysis, due to its popularity and its compatibility with our analysis tools, we selected only the repositories that contain at least 80\% of Java code. 
%Java is a widely used programming language that is compatible with the tools we employed for code similarity analysis. By focusing on Java repositories, we ensured that our examination of the methods was systematic and aligned with the capabilities of our analysis tools.

Lastly, we require that the selected repositories use Maven as a build tool. This criterion was established to ensure a consistent structure across the repositories, facilitating automation in our analysis and helping to manage the large volume of data involved. %By selecting projects built with Maven, we could streamline our processes and ensure that the repositories adhered to a common framework. 

These criteria led to the identification of 146 repositories protected by copyleft licenses. For each repository, we selected the latest version available before December 2020. %, to ensure that the analyzed code was likely processed during the training stage of \texttt{gpt-4-turbo}.

\textbf{Method Selection.} To identify the relevant methods for our analysis, we first extracted all the Java methods from the selected repository. 
%This extraction process involved gathering various metadata for each method, which were essential for subsequent steps in our analysis. The metadata collected included the project of origin, a unique identifier for the method, the file path from which the method was extracted, the method signature (comprising the method name and parameter types), the line number of the method declaration, the length of the method in terms of lines, and the length of any associated JavaDoc comments. 
This %comprehensive data collection 
resulted in the extraction of a set of 1,055,485 methods.
We excluded the test methods from the selection, resulting in a dataset of 879,337 methods. Then, we selected only the methods that meet specific criteria, motivated by the need to ensure the dataset consists of meaningful code for our analysis. Since the context of the study is the potential suggestion of code protected by copyleft licenses, only the reproduction of non-trivial methods could represent interesting and dangerous cases of unintended plagiarism. For this reason, we selected the methods with at least 10 non-empty lines of code. 
Additionally, to ensure that the documentation provides sufficient context and clarity regarding the method's purpose and usage, we required the JavaDoc of each method to have at least 100 characters and 4 non-empty lines. %To further enhance the quality of the documentation, we specified that the JavaDoc comments must exceed 100 characters, which helps to guarantee that the comments are detailed and informative rather than vague or superficial. 
These requirements led to the identification of 33,413 methods. % on which we calculated McCabe's complexity \cite{cyclomatic-complexity}.

\textbf{Method Sampling.} To sample a set of methods that can be feasibly processed from the identified methods, we randomly selected methods considering different sizes (i.e., locs) and complexities (i.e., McCabe complexity~\cite{cyclomatic-complexity}), guaranteeing that a diversity of cases occur in the final set of methods. In particular, we first computed the 33$^{rd}$ and 67$^{th}$ percentiles of the size and complexity of the selected methods. The 33$^{rd}$ and 67$^{th}$ percentiles of size are 16 locs and 27 locs, respectively. While the 33$^{rd}$ and 67$^{th}$ percentiles of complexity are 3 and 6, respectively. We then classified the methods as low/medium/high size if their size is strictly below 16 locs, between 16 and 27 locs (excluded), and above 27 locs, respectively. Similarly, we classified methods as low/medium/high complexity, if their complexity is strictly below 3, between 3 and 6 (excluded), or above 6, respectively. Table~\ref{tab:groups} shows the distribution of the 33,413 methods according to these categories.
%In this step, we ensured that the final dataset was representative and diverse. To facilitate this selection, we calculated the 33.33rd and 66.67th percentiles for various parameters, such as ad method length, JavaDoc comment length, and McCabe complexity~\cite{cyclomatic-complexity}. The results are shown in Table \ref{tab:percentiles} These percentiles provided useful thresholds for classifying methods into three categories: low, mid, and high. For instance, methods with fewer than 16 non-empty lines were classified as low, while those with more than 27 lines were considered high. Similarly, for complexity, methods with a complexity of fewer than 3 were classified as low, while those exceeding a complexity of 6 were categorized as high. After classifying the methods, the number of methods in each group was calculated, where group means the combination of categories for each parameter. The results are shown in table \ref{tab:groups}.
To select a large, but manageable, set of methods, we randomly selected 300 methods from within each group reported in Table~\ref{tab:groups}. To guarantee that a diversity of methods belonging to different source projects are selected from within each group, we divided the methods according to the project they have been extracted from, and we randomly selected a balanced set of methods from each project, based on the projects' availability. 
%This strategy ensured a good variety of methods, preventing any single group from dominating the dataset. However, it is important to note that 
For three out of the 27 groups containing less than 300 methods, we selected all the available methods. 
Thus, the final dataset %used in our experimentation 
consists of 7,347 methods.

%%%%%%%%%%%%% COMMENTATA %%%%%%%%%%%%%
\begin{comment}
\begin{table}
\centering
\begin{tabular}{c|c|c|c|}
\cline{2-4}
                                                  & \textbf{Method } & \textbf{JavaDoc } & \textbf{McCabe } \\ 
                                                 & \textbf{ Lines} & \textbf{ Characters} & \textbf{ Complexity} \\ \hline
\multicolumn{1}{|c|}{\textbf{33.33rd}} & 16                    & 188                         & 3                          \\ \hline
\multicolumn{1}{|c|}{\textbf{66.67th}} & 27                    & 325.1135                    & 6                          \\ \hline
\end{tabular}
\caption{Percentile Values} \label{tab:percentiles} \end{table}
\end{comment}
%%%%%%%%%%%%% COMMENTATA %%%%%%%%%%%%%

\begin{table}[ht]
\caption{Number of methods for each group: Method Lines (ML), McCabe Complexity (MC) and JavaDoc Characters (JC).}
   \setlength{\tabcolsep}{4pt}
\begin{tabular}{@{}llccccccccc@{}}
\toprule
\multicolumn{2}{c|}{\textbf{ML}}                                             & \multicolumn{3}{c|}{\textit{\textbf{Lo}}}                                               & \multicolumn{3}{c|}{\textit{\textbf{Me}}}                                               & \multicolumn{3}{c}{\textit{\textbf{Hi}}}                           \\ \midrule
\multicolumn{2}{c|}{\textbf{MC}}                                             & \textit{\textbf{Lo}} & \textit{\textbf{Me}} & \multicolumn{1}{c|}{\textit{\textbf{Hi}}} & \textit{\textbf{Lo}} & \textit{\textbf{Me}} & \multicolumn{1}{c|}{\textit{\textbf{Hi}}} & \textit{\textbf{Lo}} & \textit{\textbf{Me}} & \textit{\textbf{Hi}} \\ \toprule
\multicolumn{1}{c|}{}            & \multicolumn{1}{l|}{\textit{\textbf{Lo}}} & 3,407                & 1,326                & \multicolumn{1}{c|}{55}                   & 1,089                & 1,861                & \multicolumn{1}{c|}{425}                  & 378                  & 818                  & 1,893                \\ \cmidrule(l){2-11} 
\multicolumn{1}{l|}{\textbf{JC}} & \multicolumn{1}{l|}{\textit{\textbf{Me}}} & 2,928                & 1,348                & \multicolumn{1}{c|}{37}                   & 999                  & 1,795                & \multicolumn{1}{c|}{565}                  & 391                  & 902                  & 2,055                \\ \cmidrule(l){2-11} 
\multicolumn{1}{l|}{}            & \multicolumn{1}{l|}{\textit{\textbf{Hi}}} & 2,278                & 946                  & \multicolumn{1}{c|}{55}                   & 1,024                & 1,810                & \multicolumn{1}{c|}{333}                  & 370                  & 1,183                & 3,135                \\

\bottomrule
%\\
\addlinespace
\multicolumn{11}{l}{\textit{\textbf{Lo: Low, Me: Medium, Hi: High}}}                                                   
\end{tabular}
\label{tab:groups}
\end{table}

\subsection{Plagiarism Assessment}

%We evaluated the originality of the code generated by ChatGPT in response to various prompts designed around specific research questions (RQs). The assessment focuses on understanding how different prompt structures and contextual information influence the likelihood of generating code that may infringe on copyleft licenses. 

\textbf{Code generation.} %The design of prompts is a critical aspect of the code generation process, as it directly influences the responses generated by ChatGPT. 
In this study, we developed specific prompts to address each research question (RQs). The full text of the prompts is available in our dataset. %related to the potential for generating code that may infringe on copyleft licenses. Below, we detail the prompts used for each research question, highlighting their structure and intent.

To address RQ1, for each method in our dataset, we ask ChatGPT to generate a plausible implementation given the signature of the method and its JavaDoc description. %For example, we ask \texttt{\small Based on the following JavaDoc comment, generate the corresponding method: "Calculates the sum of two integers"}.\todo{Nel prompt non dovrebbe esserci anche la firma del metodo?} %This approach is specifically designed to explore whether ChatGPT generates code that resembles existing implementations when it is given only the JavaDoc comment as context. By doing so, it allows for an assessment of the model's tendency to plagiarize when provided with minimal contextual information.

To address RQ2, %which examines whether the context provided in the request influences the likelihood of returning code protected by copyleft licenses, 
we extend the prompt providing not only the signature of the method whose implementation has to be generated but also the rest of the code present in the class that embeds that method. In this way, we can study if the context may influence the likelihood possibly plagiarized code is returned by ChatGPT. %To this end, we compare the results obtained with the extended prompt to the prompt used for RQ1. 

%the prompt structure involves a clear task description that specifies the functionality required from the code. The context varies by including two different scenarios: one that presents only the JavaDoc comment followed by the method signature and another that also provides the entire class code. The instruction is to request the generation of code based on the provided context. This design facilitates a comparative analysis of how varying the context affects the likelihood of generating code that resembles existing implementations, thereby helping to understand the influence of context on the model's output.

To address RQ3, we consider a variant of the prompt used for RQ2, that is, we do not embed the full code of the class in the prompt, but we only add the access methods (i.e., getter and setter methods). This allows us to study whether the presence of common coding patterns like the presence of access methods may influence the likelihood of generating plagiarized protected code.

%which investigates whether the inclusion of access methods in the prompt affects the probability of obtaining code protected by copyleft licenses, the prompt structure begins with a task description that specifies the functionality required from the code. This structure incorporates standard access methods, such as getters and setters, that are relevant to the task at hand. The instruction is to request the generation of code that utilizes the provided access methods. This approach aims to test whether the presence of common coding patterns, such as access methods, influences the likelihood of generating plagiarized code. It seeks to explore the relationship between standard coding practices and the originality of the output.

To address RQ4, we use the same prompt used for RQ1, but we configure ChatGPT to use different temperature values. RQ1 uses the default temperature value $1$, while RQ4 investigates the effect of using the temperature values $0$ (minimum creativity) and $2$ (maximum creativity).

%In the context of RQ4, which explores whether the temperature parameter of ChatGPT influences the likelihood of returning code protected by copyleft licenses, the prompt structure begins with a JavaDoc comment that clearly states the functionality required from the code. It also specifies the desired temperature setting, which can be low, medium, or high, to control the randomness of the output. The instruction is to request the generation of code based on the specified temperature. This structure facilitates an examination of how the temperature setting affects the originality of the generated code. By comparing outputs at different temperature levels, it becomes possible to assess the model's propensity for plagiarism.

%Finally, to address RQ5, %which investigates whether ChatGPT is capable of avoiding the generation of copied code when explicitly requested, 
%we extend the prompt used for RQ1 by clearly stating that we want ChaptGPT \texttt{``to not copy any known implementation''}, expecting the original code to be returned.

Finally, to address RQ5, we modify the prompt used for RQ1 by explicitly instructing ChatGPT to  ``\texttt{not copy any known implementation}''. Our goal with this instruction is to encourage the model to generate unique code rather than returning code that might be similar to any publicly available implementation, potentially reducing the risk of generating code that infringes copyleft licenses.

%original code not obtained from known sources. This is obtained by adding the sentence . 

To account for the potential non-determinism of the responses, we collected five responses for each request, and we used only the methods that produced valid responses in the study. In particular, for RQ1 and RQ2 we used 6,917 methods out of 7,347. For RQ3, we restricted the analysis to the methods that were reported as likely plagiarized with RQ2, which considers the class of the method as a context, and studied if they were still likely plagiarized when using the access methods only as context. We thus used 631 valid methods out of 675 selected methods. For RQ4 and RQ5, we restricted the analysis to the methods that were likely plagiarized according to RQ1 (only the method signature and JavaDoc comment used in the prompt), finally using 233 out of 239 selected methods. The invalid responses that slightly reduced the set of methods used in the experiment were due to ChatGPT returning incorrect signatures, truncated responses, or syntactic errors that prevented the automatic analysis of the response. Overall, to answer these research questions, we collected more than 70K method implementations.

\begin{comment}
\begin{itemize}
    \item \textbf{RQ1}: Out of 7,347 requests made, 6,917 methods were obtained for which it is possible to analyze the results.
    \item \textbf{RQ2}: Out of 7,347 requests made, 6,917 methods were obtained for which it is possible to analyze the results.
    \item \textbf{RQ3}: Out of 675 requests made, 631 methods were obtained for which it is possible to analyze the results.
    \item \textbf{RQ4}: Out of 7,347 requests made, 6,917 methods were obtained with temperature 0 for which it is possible to analyze the results; out of 239 requests made, 233 methods were obtained with temperature 2 for which it is possible to analyze the results.
    \item \textbf{RQ5}: Out of 239 requests made, 235 methods were obtained for which it is possible to analyze the results.
\end{itemize}
\end{comment}

%unsuccessful requests can be attributed to various reasons,which prevent automatic analysis of the generated code. Some errors were caused by incorrect signatures, such as the addition of unnecessary keywords or the replacement of existing keywords. Other issues arose from comments within the signature, which led to a mismatch between the original signature and the one returned by ChatGPT. Additionally, some responses were truncated due to exceeding the token limit, while syntactic errors, such as incorrect characters or consecutive periods, further complicated the code generation. Finally, in some cases, the generated code required additional completion by the end user, rendering the response incomplete.

\textbf{Plagiarism detection.} %We then analyzed the results obtained from the code generation requests submitted to ChatGPT, focusing on the metrics used to compute similarity and detect possible plagiarism.
We use two main metrics to assess the similarity between the code returned by ChatGPT and the method implementation protected by the copyleft license available in our dataset.

%\begin{itemize}[leftmargin=*]
    %\item 
    \emph{Similarity}: This metric represents the level of similarity between two methods that are compared. We use  JPlag~\cite{jplag}, a popular plagiarism detection tool, to quantify the similarity between two implementations. For this study, we employed the default configuration of JPlag, which does not incorporate token normalization, and we used JPlag's average similarity function as our similarity metric. JPlag measures similarity as the proportion of matching tokens that can be identified in the compared code. It is used to estimate with an index ranging between $0$ (totally dissimilar) and $1$ (totally similar) how likely an implementation is a plagiarized version of another implementation while abstracting from irrelevant syntactic changes. We use this metric to determine how likely the code recommended by ChatGPT can be considered the plagiarized version of the copyleft code.  In particular, for each method $m_{orig}$ used in the study, given a set of responses $m_i$ obtained by submitting the same request multiple times, we computed both the \emph{mean} and \emph{max similarity}, defined as the mean and max value of  $\{ \textit{similarity}(m_{orig}, m_i) \}$, respectively. The mean value represents the average case, while the max value represents the worst case, in terms of the plagiarized code that might be recommended to a developer. 
  %  That is, it measures  a comprehensive view of how closely related the two pieces of code are, taking into account all the tokens present in each submission.  The Average Similarity score ranges from 0 to 1, with higher values indicating greater similarity. This metric is particularly valuable for identifying potential cases of plagiarism, as it allows for an in-depth assessment of the extent of similarity between different code samples. By calculating average similarity, one can gain insights into the overall relationship between various pieces of code, facilitating a more informed analysis of potential intellectual property violations.
    
    %\item 
    \emph{Fuzzy Ratio}: We complement the similarity as computed by JPlag, which works at the token level, with a metric working at the character-level, that is more sensitive to irrelevant syntactic changes, but, for this same reason, better capturing \emph{nearly identical} copies of code. To this end, we used the fuzzy ratio, which is defined as one minus the ratio between the Levenshtein distance of the compared strings, computed as the single-character edits (insertions, deletions, or substitutions), and the maximum length of the strings. We used TheFuzz~\cite{fuzzy} to make this computation. The resulting value is between 0 and 1, with 1 indicating equals strings and $0$ totally dissimilar strings. Also for the fuzzy ratio, we compute the \emph{mean} and \emph{max fuzzy ratio}, as done for the similarity, to address the fact the same requests are repeated multiple times. 
    
    %    purely syntactic metric This is a syntactic similarity measure, calculated by TheFuzz \cite{fuzzy}, that captures only the syntactic differences between two pieces of code. It is calculated using the Levenshtein distance, which measures how many single-character edits (insertions, deletions, or substitutions) are required to change one string into another. We expressed the fuzzy ratio on a scale from 0 to 1, where a higher value indicates greater similarity. It is an effective metric for detecting minor variations in code that may suggest copying.
%\end{itemize}

%To analyze whether the various responses collected from ChatGPT for each type of request generate equivalent similarity values, we calculated and verified the variance among the values for average similarity and fuzzy ratio. Low variance values indicate that the data related to the various responses can be summarized with a single value, which approximates all responses due to the low variance. This approximation can be achieved by considering the average value that reflects the model's average tendency regarding plagiarism. It is important to note that low variance does not imply that the responses are identical, but rather that they have a similarity value relative to the original code that is equivalent or similar.

\smallskip 

To determine if plagiarism has occurred, it is necessary to establish a similarity threshold above which cases are considered suspicious. It is hard to establish an exact threshold, but any code reuse to represent the outcome of a potentially illegitimate action must have a lot in common with the source. For these reasons, rather than using a single hard threshold, we discuss results for similarity values above $0.7$, considering every possible threshold value with a step of $0.05$. To confirm our design choice to focus on cases above $0.7$, we inspected a selection of the responses provided by ChatGPT with similarity values spanning from $0.45$ to $0.95$. We could confirm that the vast majority of the cases with similarity above $0.7$ are indeed suspicious, while only a few cases with similarity below $0.7$ are suspicious. 
%This value of the threshold has also been used in similar studies \cite{}\todo{add cit}, corroborating our choice.

%To estimate this threshold, we manually inspected 36 methods  randomly selected responses and classified them as being likely the result of plagiarism or not. The ... \todo{add percentage} of the results with a level of similarity above $0.7$ were classified as plagiarism, while for lower levels the occurrences of possible plagiarism were much more uncertain and rare. We thus used this threshold when reporting the results. This value of the threshold has also been used in similar studies \cite{}\todo{add cit}, corroborating our choice. However, since this is not an exact threshold but an estimated threshold, we will report also results for higher values of these threshold, to account for the sensitivity of the results to the choice of the threshold.  

%This threshold is conservative and aims to classify only actual cases of suspected plagiarism, reducing the likelihood of false positives while increasing the risk of false negatives. It has been observed that methods with a similarity above 0.7 can be considered cases of suspected plagiarism \cite{}. However, even below this threshold, there may still be instances of plagiarism, albeit less frequently.

\textbf{Analysis of the Results.} We conducted a statistical analysis to assess the influence of various factors on the likelihood of obtaining code that resembles copyleft code. As factors, we studied the influence of the context (RQ2 and RQ3), the temperature (RQ4), and the prompt (RQ5).

To answer RQ1, we compute the mean and maximum similarity and fuzzy scores, for the whole set of methods in the dataset. 
To answer the rest of the research questions (RQ2-RQ5), we compute the differences between the (mean and maximum) similarity and fuzzy scores of corresponding methods whose code is obtained by changing one factor in the request. That is, for each score (mean similarity, max similarity, mean fuzzy ratio, and max fuzzy ratio), we compute for each method $m$ the value $\textit{score}(m)_{\textit{base-request}}-\textit{score}(s)_{\textit{changed-factor}}$, where \textit{base-request} is the request as formulated for RQ1, and the \textit{changed-factor} is the same request with a factor modified according to the research question (RQ2 and RQ3 use a different context, RQ4 a different temperature value, and RQ5 a different prompt). 
A value of 0 indicates that the compared scores are identical for the considered method, while a negative (positive) value indicates that the case with the changed factor leads to code that has a higher (lower) similarity to the copyleft code. The distribution of these differences is indicative of how the studied factor influences the results on a per-method basis.

To further study the impact of each factor, we also compare the population of all score values (mean similarity, max similarity, mean fuzzy ratio, and max fuzzy ratio) for the code obtained with the \textit{base-request} and the \textit{changed-factor}.
We check for the presence of any statistically significant difference ($\alpha = 0.05$) between the compared populations using the Wilcoxon signed-rank test~\cite{test-wilcoxon}, and we compute the effect size $r$ of the significant differences~\cite{cohen2013statistical}, to discover any very small ($ |r| < 0.1 $), small ($ 0.1 \leq |r| < 0.3 $),  medium ($ 0.3 \leq  |r| < 0.5 $),  and large ($ |r| \geq 0.5 $) effect, which is indicative of the practical significance of the observed phenomena.

\section{Results}
\label{sec:Res}
This section presents the results obtained by evaluating the code generated by ChatGPT in response to various prompts designed to explore the potential for generating code covered by copyleft licenses. %The presentation is structured around the five RQs outlined in the methodology.

\subsection{RQ1: Can ChatGPT return code protected by copyleft licenses?}

To answer this research question, we prompted ChatGPT for the implementation of the methods in our dataset starting from the method's signature and its JavaDoc comment, which served as a minimal directive for the model. We maintained the temperature to its default value (i.e., $1$). 

\begin{figure} \centering \includegraphics[width=0.4\textwidth]{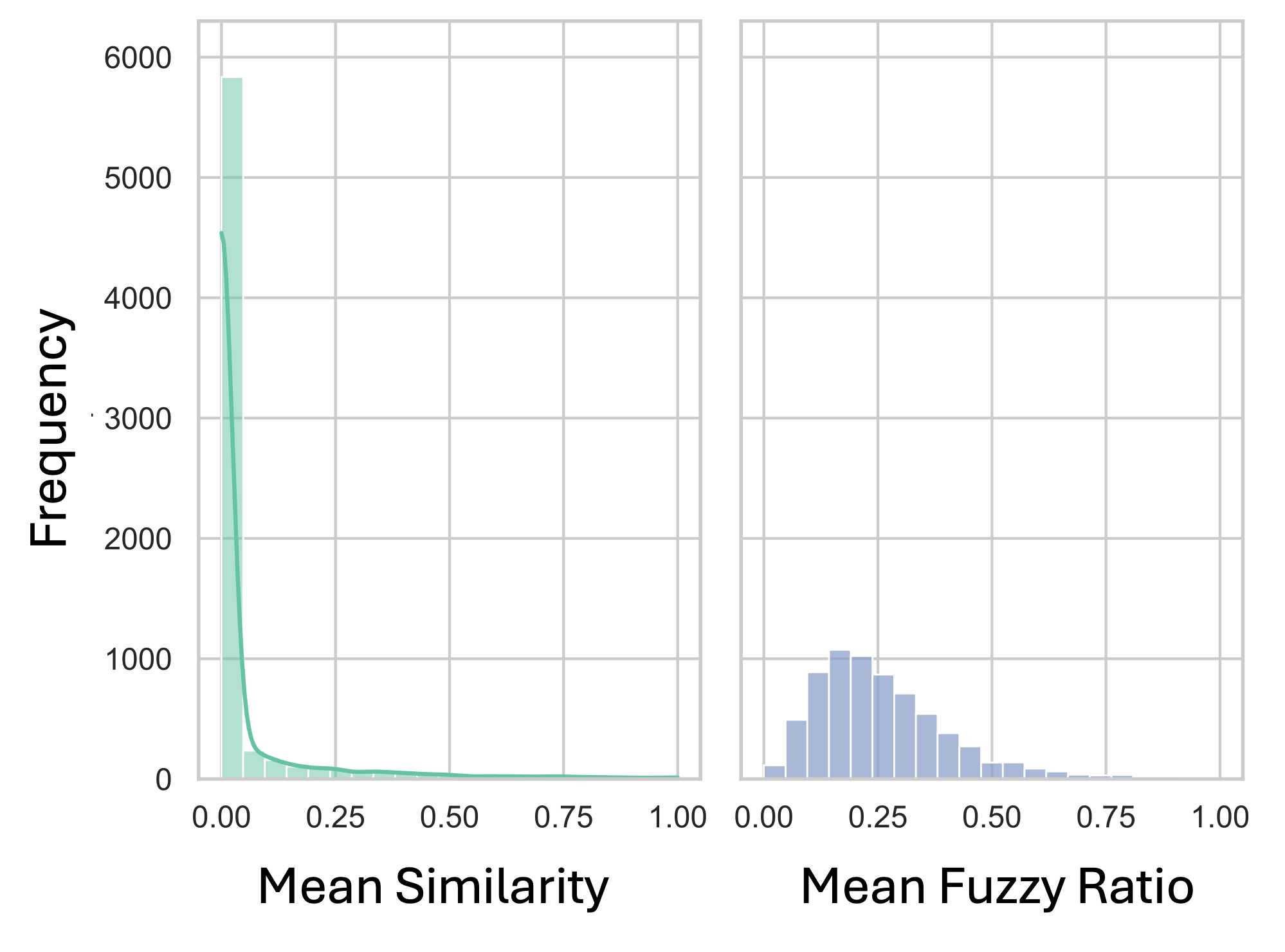} 
\vspace{-10pt} 

\caption{Distribution of similarity scores for \emph{mean similarity}.} \label{fig:similarity_distribution} 
\vspace{-5pt} 
\end{figure}

\begin{figure} 
\centering 
\includegraphics[width=0.4\textwidth]{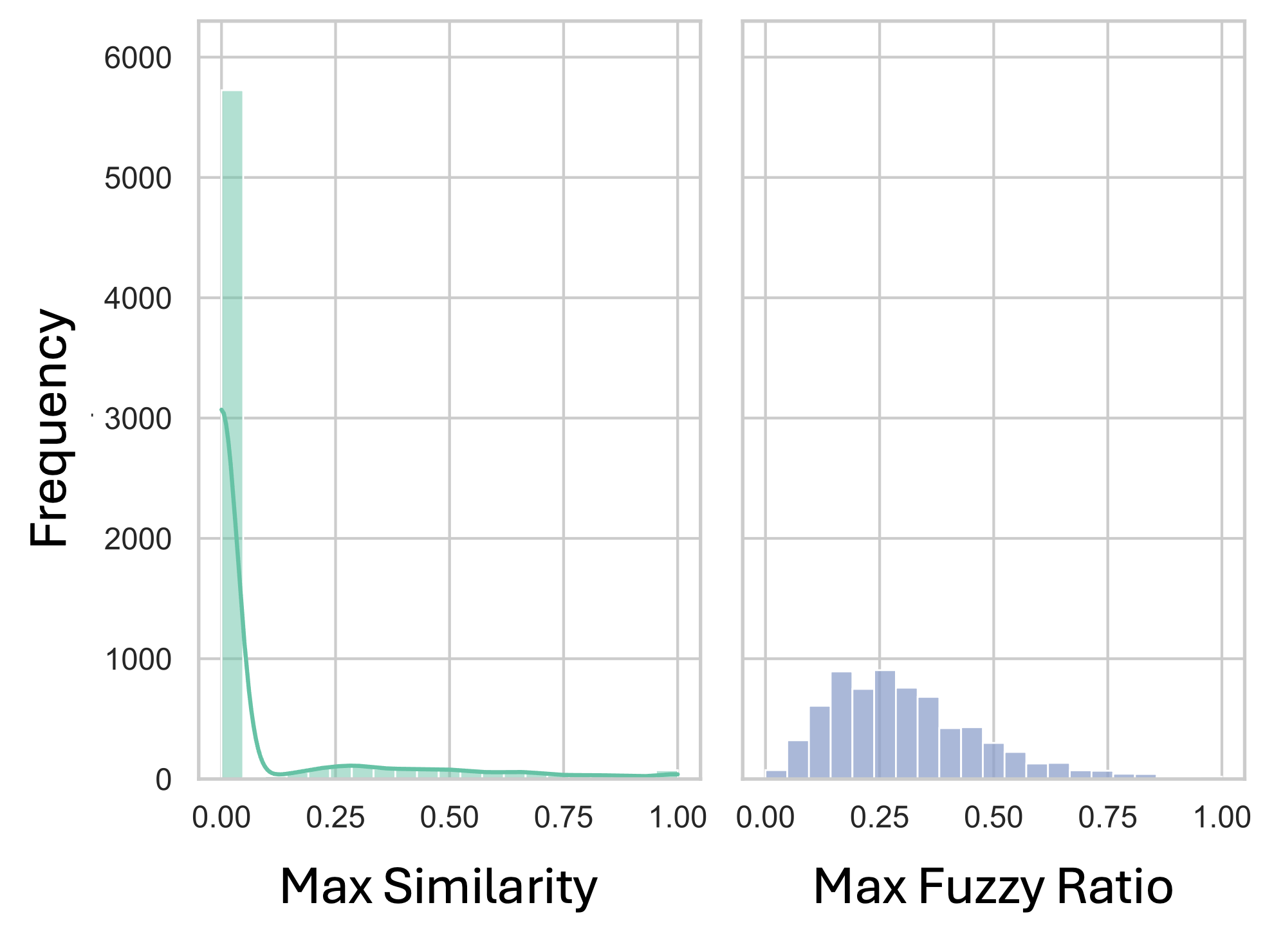} 
\vspace{-10pt} 

\caption{Distribution of similarity scores for \emph{max similarity}.} 
\label{fig:max_similarity_distribution} 
\vspace{-5pt} 
\end{figure}

As illustrated in Figures~\ref{fig:similarity_distribution} and~\ref{fig:max_similarity_distribution}, the analysis of the responses reveals that the significant majority of the obtained implementations (i.e., 82.67\% for mean and max similarities) have a similarity score equals to $0$. This indicates that the generated code is significantly different from the copyleft code available in the dataset, and only occasionally the generated code closely mirrors existing implementations.

%In the case of maximum similarity shown in Figure \ref{fig:max_similarity_distribution}, we observe that, as expected due to the use of higher similarity values, a greater number of methods exhibit higher similarity scores compared to the average similarity case. Notably, there are more instances with a similarity score of 1. This suggests that while most generated outputs are distinct, some closely mirror existing implementations, raising concerns about potential copyright infringement. Despite this increase in similarity scores, the overall distribution pattern remains consistent with that of the mean similarity.

The fuzzy ratio measuring syntactic similarity is slightly higher than mean and max similarities because some common syntactic elements could be found even between rather diverse implementations. Yet the vast majority of the code has a low fuzzy ratio.% it is more sensitive to small syntactic changes, which can lead to significant differences in its values even when the underlying logic remains similar.\todo{Credo sia l'opposto, il grafico mostra maggiore similarità ma il commento dice la cosa opposta}

We zoom into the cases with suspicious similarity levels reporting in Table~\ref{tab:count_methods_similarity} the percentage of methods with a similarity above $0.7$. Specifically, 102 methods (1.48\%) have a mean similarity above 0.70, while 232 methods (3.35\%) exceed this threshold for maximum similarity. These results indicate that, although infrequent, some responses warrant caution in reuse. 

For instance, consider the method from our dataset presented in Listing~\ref{code}. Creating an implementation based solely on its JavaDoc comment and signature poses significant challenges. However, ChatGPT managed to generate code, as shown in Listing~\ref{generatedCode}, that closely resembles the original, achieving a similarity score of $1.0$ and a fuzzy ratio of $0.86$. This generated code incorporates only minor modifications that enhance readability while preserving the functionality, even delivering the same output messages through \texttt{System.out} statements.

\noindent \fbox{\parbox{0.98\columnwidth}{\textbf{Answer to RQ1} ChatGPT seldom generates code potentially violating copyleft licenses (3.35\% of the cases in the worst case). Yet, users cannot entirely ignore the risk of incidentally using the copyleft code.}}

\begin{table}[]
    \centering
\caption{Count of methods for mean similarity values.} 
\begin{tabular}{c|c|c}
\toprule
\textbf{Similarity}             & \textbf{Mean Similarity} & \textbf{Max Similarity} \\
\textbf{Values}                 & \textbf{Frequency}       & \textbf{Frequency}      \\ \midrule
1.00                            & 11 (0.16\%)              & 61 (0.88\%)             \\
{[}0.95, 1.00{]}                & 8 (0.12\%)               & 12 (0.17\%)             \\
{[}0.90, 0.95{]}                & 11 (0.16\%)              & 23 (0.33\%)             \\
{[}0.85, 0.90{]}                & 11 (0.16\%)              & 29 (0.42\%)             \\
{[}0.80, 0.85{]}                & 19 (0.27\%)              & 36 (0.52\%)             \\
{[}0.75, 0.80{]}                & 13 (0.19\%)              & 34 (0.49\%)             \\
{[}0.70, 0.75{]}                & 29 (0.42\%)              & 37 (0.53\%)             \\ \midrule
\textbf{Total {[}0.70, 1.00{]}} & \textbf{102 (1.48\%)}    & \textbf{232 (3.35\%)}   \\ \bottomrule
\end{tabular}
%\vspace*{1mm}
\label{tab:count_methods_similarity}
\end{table}

%\begin{lstlisting}[language=Java,caption={An Example of a Method from Our Dataset.}, basicstyle=\small\ttfamily,breaklines=true,columns=fullflexible, frame=single, label=code]

\noindent
\begin{minipage}{1\linewidth} 
\begin{lstlisting} [label=code,frame=single,caption={An example of a method from our dataset.}]
/**
 * To solve this question, we just need to count the
 * number of persons that overtake a particular 
 * person.
 *
 * @param q the queue
 */
private static void minimumBribes(int[] q) {
  int bribes = 0;
  for (int i = q.length - 1; i >= 0; i--) {
    if (q[i] - i - 1 > 2) {
      System.out.println("Too chaotic");
      return;
    }
    for (int j = Math.max(0, q[i] - 2); j < i; j++) {
      if (q[j] > q[i]) bribes++;
    }
  }
  System.out.println(bribes);
}   
\end{lstlisting} 
\end{minipage}
\noindent
%[basicstyle=\small\ttfamily,breaklines=true,columns=fullflexible, frame=single, label=generatedCode,caption={Method Generated by ChatGPT.}]
\noindent
\begin{minipage}{1\linewidth} 

\begin{lstlisting} [label=generatedCode,frame=single,caption={Method generated by ChatGPT.}]
private static void minimumBribes(int[] q) {
  int bribes = 0;
  for (int i = 0; i < q.length; i++) {
    if (q[i] - (i + 1) > 2) {
      System.out.println("Too chaotic");
      return;
    }
    for (int j = Math.max(0, q[i] - 2); 
                                    j < i; j++) {
      if (q[j] > q[i]) {
        bribes++;
      }
    }
  }
  System.out.println(bribes);
}   
\end{lstlisting} \end{minipage}
\noindent

\subsection{RQ2: Does the context provided in the request influence the likelihood of returning code protected by copyleft licenses?}

RQ2 investigates whether the context provided in the requests to ChatGPT affects the likelihood of generating code protected by copyleft licenses. %The context can significantly shape the output, which may guide the model towards more specific responses.
To analyze the influence of context on the code generation capabilities of ChatGPT, we compared results obtained with minimal information (i.e., the method signature and the JavaDoc comment already used in RQ1) to an extensive context that includes the code present in the class with the method. The goal is to determine if a broader context increases the chances of generating code similar to the existing code. 

For example, consider the method \texttt{getSpeakerBytes()} of the CoreNLP project present in our dataset. When we prompted ChatGPT using only the method's signature and JavaDoc comment, the generated implementation exhibited substantial differences compared to the existing copyleft code. 
However, when we provided the rest of the class as context, ChatGPT produced variants of this method that closely mirrored the original implementation. This resulted in a similarity score of $1$, raising significant concerns regarding potential intellectual property infringement.

\begin{figure}[h]
    \centering
    \includegraphics[width=0.4\textwidth]{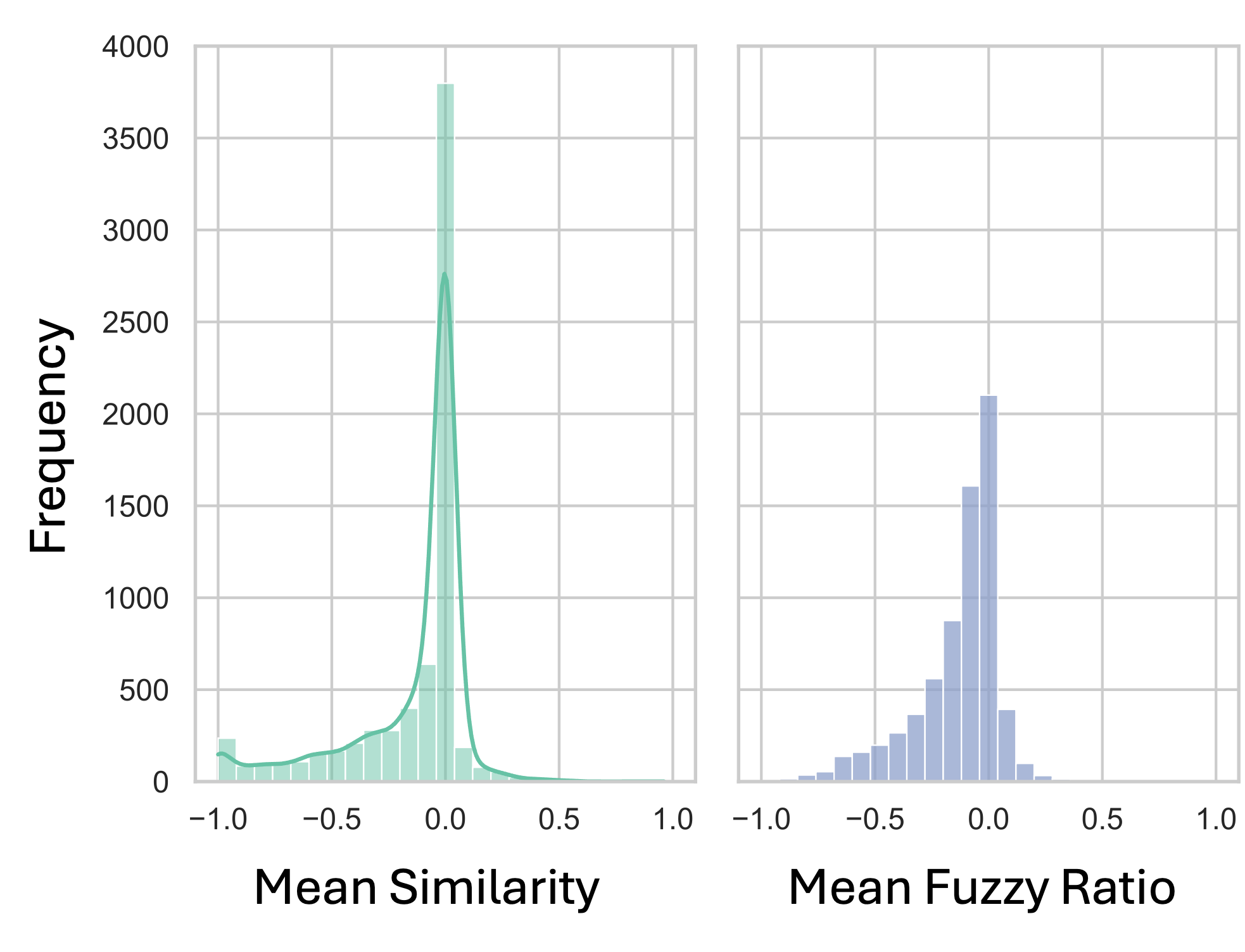}
    \vspace{-10pt} 
    
    \caption{Difference between JavaDoc comment and whole class in the case of mean similarity.}
    \label{fig:similarity_differences_in_mean_similarity}
    \vspace{-5pt} 
\end{figure}

\begin{figure}[h]
    \centering
    \includegraphics[width=0.4\textwidth]{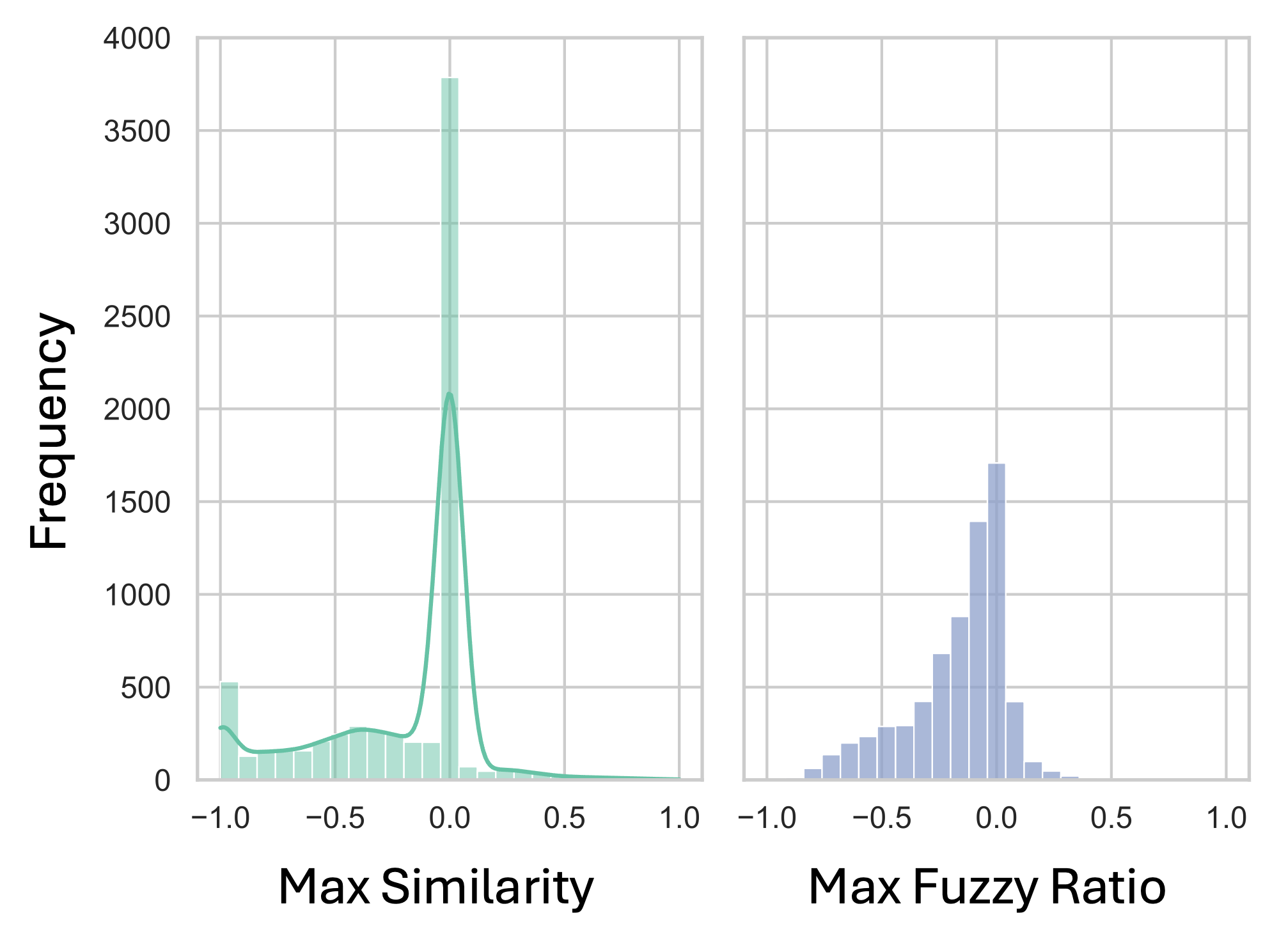}
    \vspace{-10pt} 
    
    \caption{Difference between JavaDoc comment and whole class in the case of max similarity.}
    \label{fig:similarity_differences_in_max_similarity}
    \vspace{-5pt} 
\end{figure}

Figure~\ref{fig:similarity_differences_in_mean_similarity} (left side) shows the distribution of the difference between the mean similarity obtained with minimal context (i.e., methods signature and JavaDoc) and the mean similarity obtained with the class context (i.e., all the code in the class except the body of the method that has to be generated) for each method. Figure~\ref{fig:similarity_differences_in_mean_similarity} (right side) reports the same information for the mean fuzzy ratio.  Analogously, Figure~\ref{fig:similarity_differences_in_max_similarity} shows the same differences when max similarity and max fuzzy ratio are considered.%\todo{forse sarebbe preferibile mettere nelle figure sempre mean e max prima di similarity e fuzzy ratio}

We can observe a general prevalence of cases with differences equal to $0$, that is, the mean similarity and the fuzzy ratio of the generated code are the same for both contexts. On the other hand, all the distributions are skewed on the left, that is, the context affects the similarity for several cases, increasing the chance of generating code similar to already existing copyleft code.


\begin{figure}[ht] \centering \includegraphics[width=0.4\textwidth]{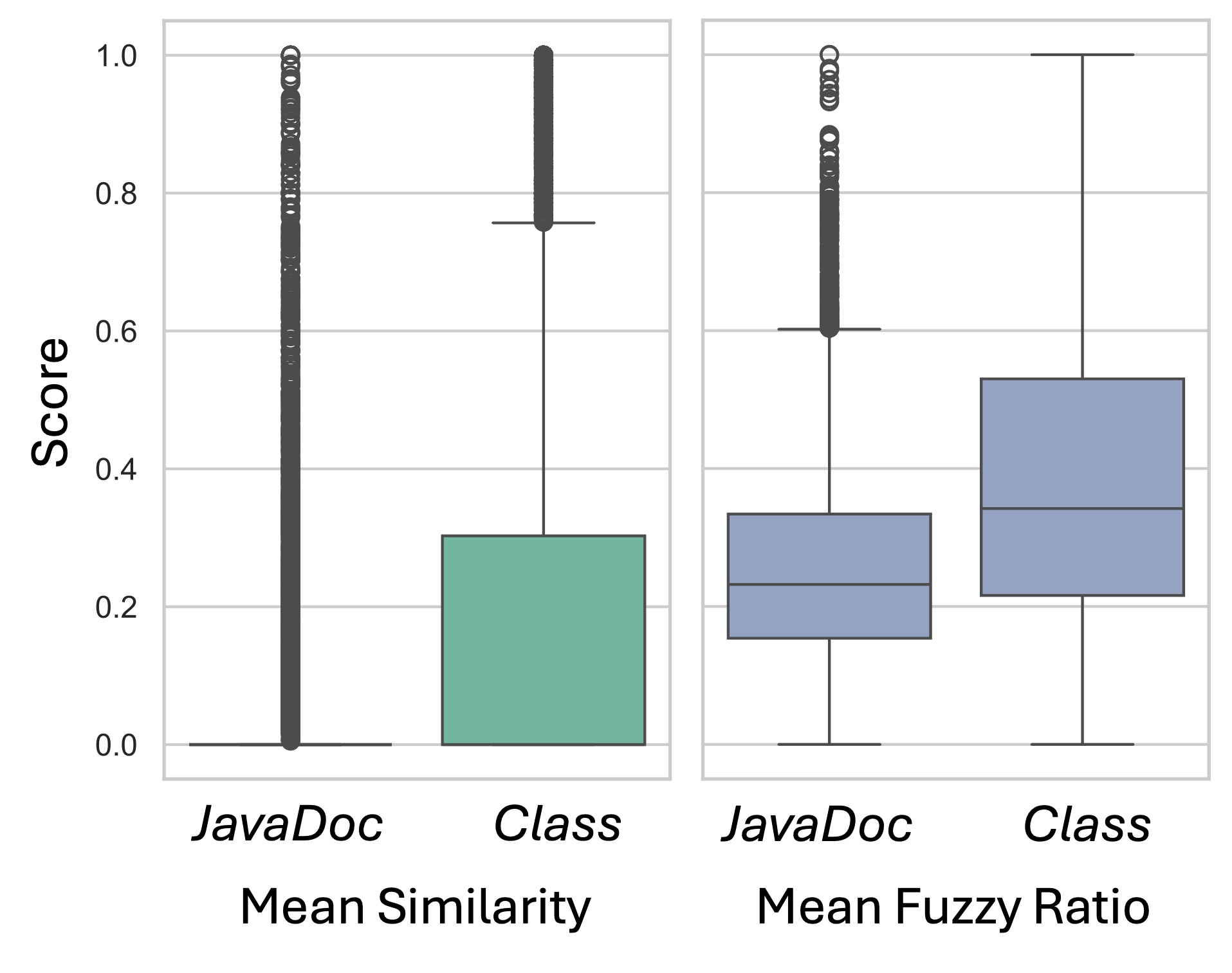} 
\vspace{-10pt} 

\caption{Similarity Scores for JavaDoc comment and whole class contexts in the case of mean similarity.} 
\vspace{-5pt} 

\label{fig:similarity_javadoc_vs_file_in_mean_similarity} \end{figure}

\begin{figure}[ht] \centering \includegraphics[width=0.4\textwidth]{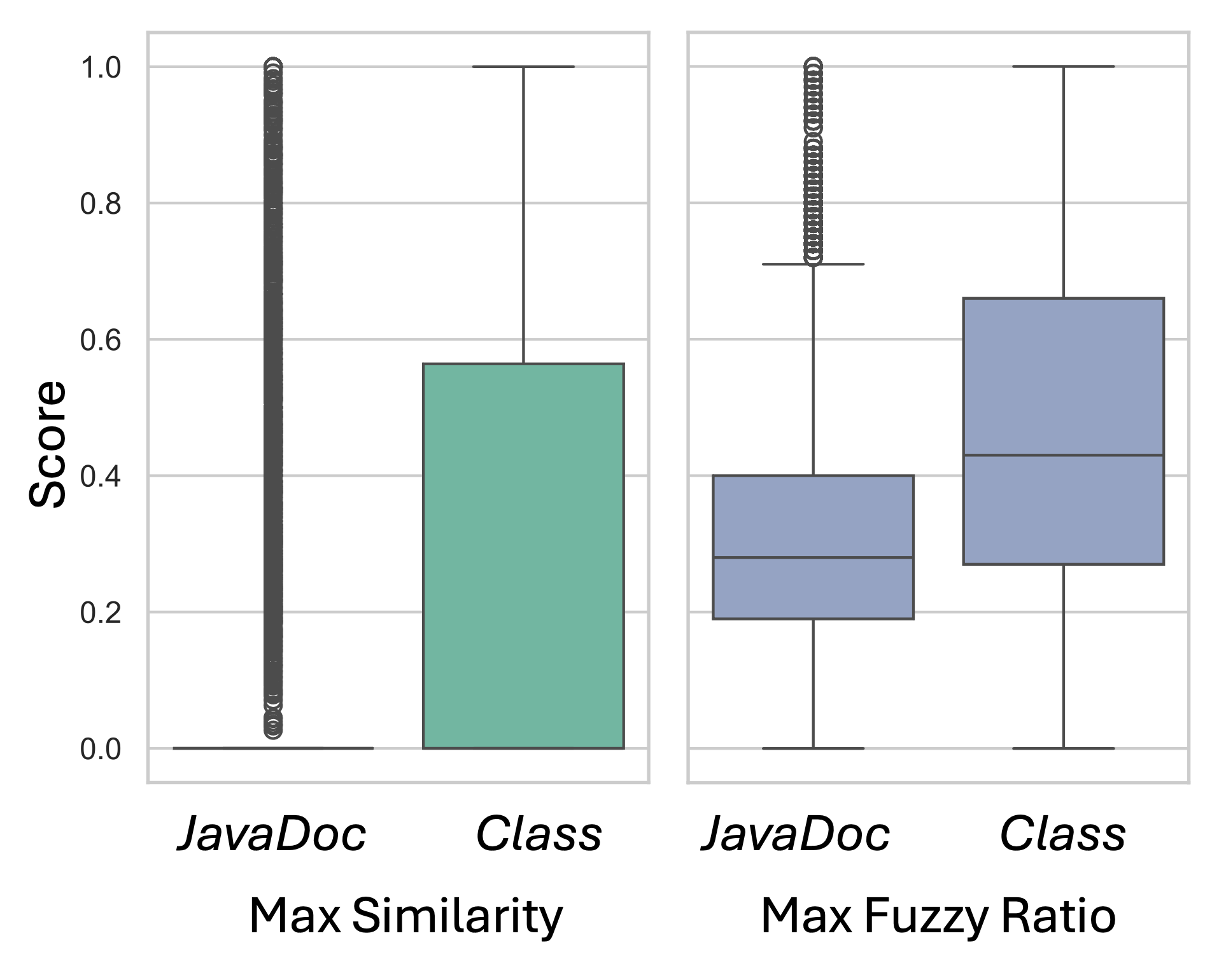} 
\vspace{-10pt} 

\caption{Similarity Scores for JavaDoc comment and whole class contexts in the case of max similarity.} 
%\vspace{-5pt} 

\label{fig:similarity_javadoc_vs_file_in_max_similarity} \end{figure}

Figures~\ref{fig:similarity_javadoc_vs_file_in_mean_similarity} and~\ref{fig:similarity_javadoc_vs_file_in_max_similarity} present the distribution of the mean and max similarity values, as well as mean and max fuzzy ratio, respectively. The boxplots indeed show that methods generated with the broader context exhibit higher similarity and fuzzy ratio values than those generated with only the JavaDoc comment. 
A Wilcoxon test ($\alpha = 0.05$) confirmed differences are significant for mean similarity, max similarity, mean fuzzy ratio, and max fuzzy ratio.  The effect sizes of the mean and max similarities are -0.7330 and -0.7434, respectively, indicating that the context is not only significant but also introduces a \emph{large} effect on similarity values. 

The strong effect sizes across all metrics indicate that the context provided to ChatGPT plays a crucial role in determining the similarity of the generated code with the already existing one. %This further supports the hypothesis that a broader context not only increases the likelihood of generating code that closely resembles existing implementations but also enhances the overall similarity metrics. The findings highlight the importance of context in guiding the output of AI models like ChatGPT, particularly in scenarios where the risk of plagiarism is a concern.
%In conclusion, the analysis demonstrates that providing a more comprehensive context significantly influences the likelihood of generating code that mirrors existing works, thereby increasing the potential for plagiarism. The statistical significance and strong effect sizes observed in the results underscore the necessity for careful consideration of the context when utilizing AI for code generation.

We also examined how many responses achieve a \textit{mean similarity} and a \textit{max similarity} above the threshold of $0.70$. The results, shown in Table~\ref{tab:count_methods_similarity_file}, indicate a significant increase in the number of methods across all ranges compared to those reported with the context limited to the JavaDoc comment. Notably, the percentage of potentially plagiarized methods increases from 1.48\% to 9.73\% (6.5X increase) in the case of \textit{mean similarity}. This percentage raises from 3.35\% to 19.50\% (5.8X increase) in the case of \textit{max similarity}. In both scenarios, the total number of potential plagiarisms is more than quintupled, reaching significant percentages.

\begin{table}[ht] \begin{center}
\caption{Count of methods for similarity values over threshold.}
\begin{tabular}{c|c|c}
\toprule
\textbf{Similarity}             & \textbf{Mean Similarity} & \textbf{Max Similarity} \\
\textbf{Values}                 & \textbf{Frequency}       & \textbf{Frequency}      \\ \midrule
1.00                            & 214 (3.09\%)             & 612 (8.85\%)            \\
{[}0.95, 1.00{]}                & 72 (1.04\%)              & 94 (1.36\%)             \\
{[}0.90, 0.95{]}                & 74 (1.07\%)              & 84 (1.21\%)             \\
{[}0.85, 0.90{]}                & 66 (0.95\%)              & 122 (1.76\%)            \\
{[}0.80, 0.85{]}                & 90 (1.30\%)              & 159 (2.30\%)            \\
{[}0.75, 0.80{]}                & 77 (1.11\%)              & 137 (1.98\%)            \\
{[}0.70, 0.75{]}                & 80 (1.16\%)              & 141 (2.04\%)            \\ \midrule
\textbf{Total {[}0.70, 1.00{]}} & \textbf{673 (9.73\%)}    & \textbf{1349 (19.50\%)} \\ \bottomrule
\end{tabular}
%\vspace*{1mm}

 \label{tab:count_methods_similarity_file}
\end{center}\end{table}

\noindent \fbox{\parbox{0.98\columnwidth}{\textbf{Answer to RQ2} If the class context is part of the prompt, the probability that the recommended method implementation resembles copyleft code increases by a factor greater than $5X$. This suggests that inadvertently accepting code recommendations that mirror already existing code may increase the chance of receiving additional recommendations of plagiarized code in the future, exposing developers to the risk of accumulating a non-trivial amount of unwanted copyleft code in their implementation.}}

 \subsection{RQ3: Does providing a context with only access methods affect the likelihood of obtaining code protected by copyleft licenses?}

Since RQ2 determined that the context is a significant factor, RQ3 investigates if contexts smaller than the entire code in the class may also affect the results. In particular, it considers the case where the context carries little information, just retaining some syntactic similarity with an existing copyleft class. Specifically, it investigates if including the same \emph{access methods} (i.e., getter and setter methods) present in an existing copyleft class may alter the likelihood of receiving a recommendation with a possibly plagiarized method from that same copyleft class.
%that are not highly characteristic of the class from which the method is extracted can incentivize ChatGPT to produce plagiarized content. The body of the message includes the access methods present in the code file from which the method was extracted.

To examine the impact of access methods, we compare the similarity values obtained by using two distinct contexts in the prompt: one case uses the JavaDoc comment and the method's signature (see RQ1), while the other case adds the access methods in the class to the context. %By setting the temperature parameter to 1, we aimed to maintain a high level of variability in the generated outputs while isolating the impact of the context provided.

\begin{figure}[h]
    \centering
    \includegraphics[width=0.4\textwidth]{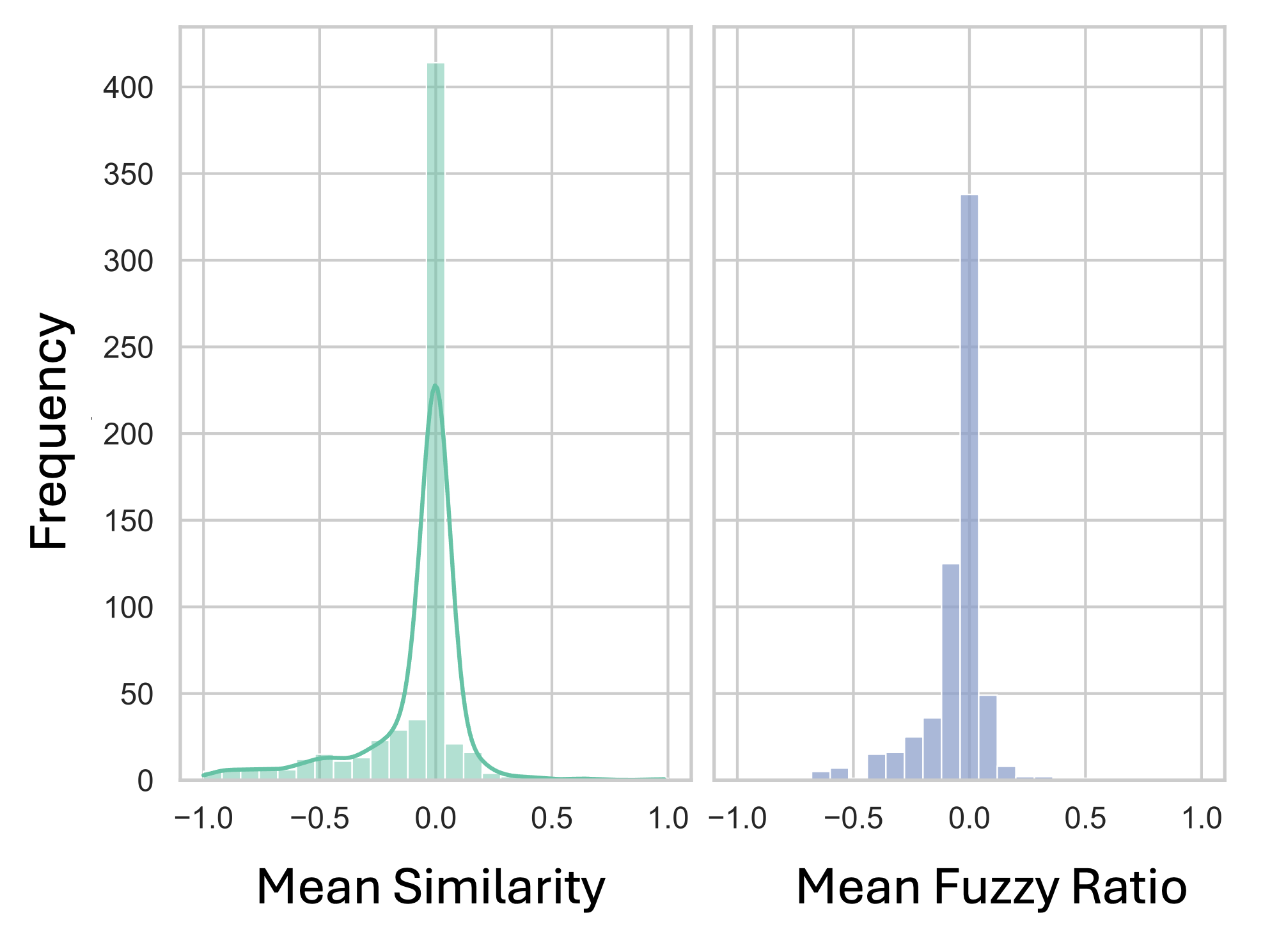}
    \vspace{-10pt} 
    
    \caption{Differences in similarity for cases of mean similarity.}
    \label{fig:similarity_differences_acess_in_mean_similarity}
\end{figure}

\begin{figure}
    \centering
    \includegraphics[width=0.4\textwidth]{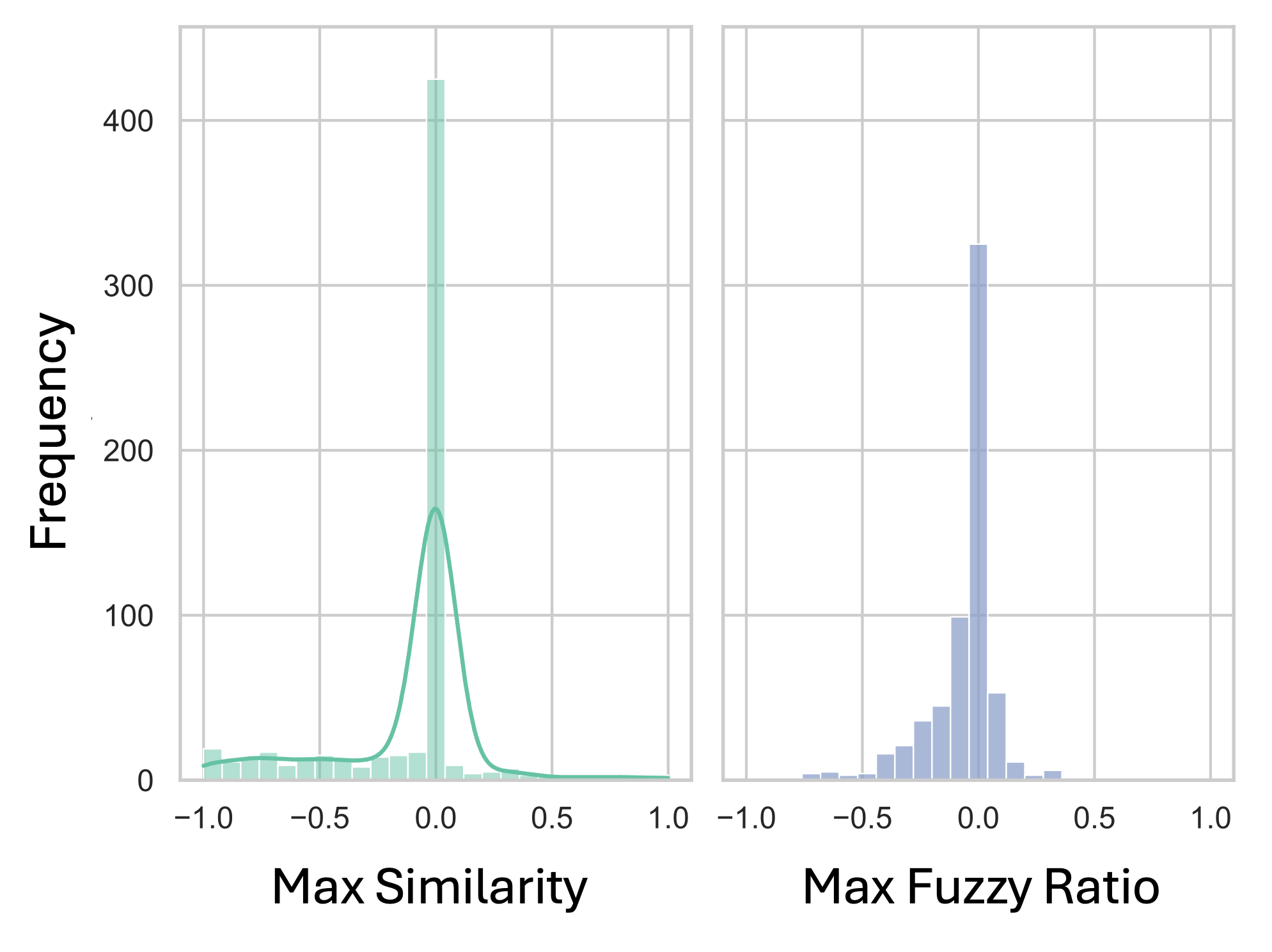}
    \vspace{-10pt} 
    
    \caption{Differences in similarity for cases of max similarity.}
    \label{fig:similarity_differences_acess_in_max_similarity}
\end{figure}

Figure~\ref{fig:similarity_differences_acess_in_mean_similarity} (left side) shows the distribution of the differences between the mean similarity values obtained with minimal context (i.e., methods signature and JavaDoc) and the mean similarity obtained with access methods.  Figure~\ref{fig:similarity_differences_acess_in_mean_similarity} (right side) reports the same information for the mean fuzzy ratio. Analogously, Figure~\ref{fig:similarity_differences_acess_in_max_similarity} shows the same differences when max similarity and max ratio are considered.

While several methods show no difference in similarity and fuzzy ratio values, distribution is skewed on the left, that is, the presence of access methods affects similarity, increasing the chance of generating code similar to existing copyleft code.

%The analysis revealed that providing also access methods as context results in higher similarity scores compared to using only the JavaDoc comment. This suggests that a more comprehensive context allows ChatGPT to access a richer set of information, thereby increasing the chances of generating code that closely aligns with existing implementations.

%Figure \ref{fig:similarity_differences_acess_in_mean_similarity} illustrates the differences in similarity values for mean similarity cases. It shows a high number of methods with a similarity difference of zero and a significant amount of negative values, indicating that including access methods as context leads to greater similarity compared to using only the JavaDoc comment. The distribution skews left, with many values near zero, suggesting that methods generated with a broader context are more closely aligned with the originals. In contrast, the right side of the graph displays few positive values, where the JavaDoc comment context results in greater similarity, but these instances are rare and also close to zero.

The plots of mean similarity and fuzzy values (Figure~\ref{fig:similarity_differences_acess_in_mean_similarity}), as well as the plot of max similarity and fuzzy values (Figure~\ref{fig:similarity_differences_acess_in_max_similarity}), show a significant impact of access methods on the level of similarity of the recommended code to the copyleft code.

%When analyzing max similarity, the results obtained by selecting the highest similarity values are visualized in Figure \ref{fig:similarity_differences_acess_in_max_similarity}. Unlike the mean similarity case, where many methods had differences close to zero, the max similarity scenario shows a decrease in the number of methods with differences close to zero. Consequently, there is an increase in the number of methods with higher similarity values.
%Figure \ref{fig:similarity_javadoc_vs_access_in_max_similarity} reveals that the max similarity values for the access methods show a significant upward shift from the mean similarity case. This confirms, as shown for RQ2, that the choice of context has a profound impact on similarity scores.

\begin{figure}[h] \centering \includegraphics[width=0.4\textwidth]{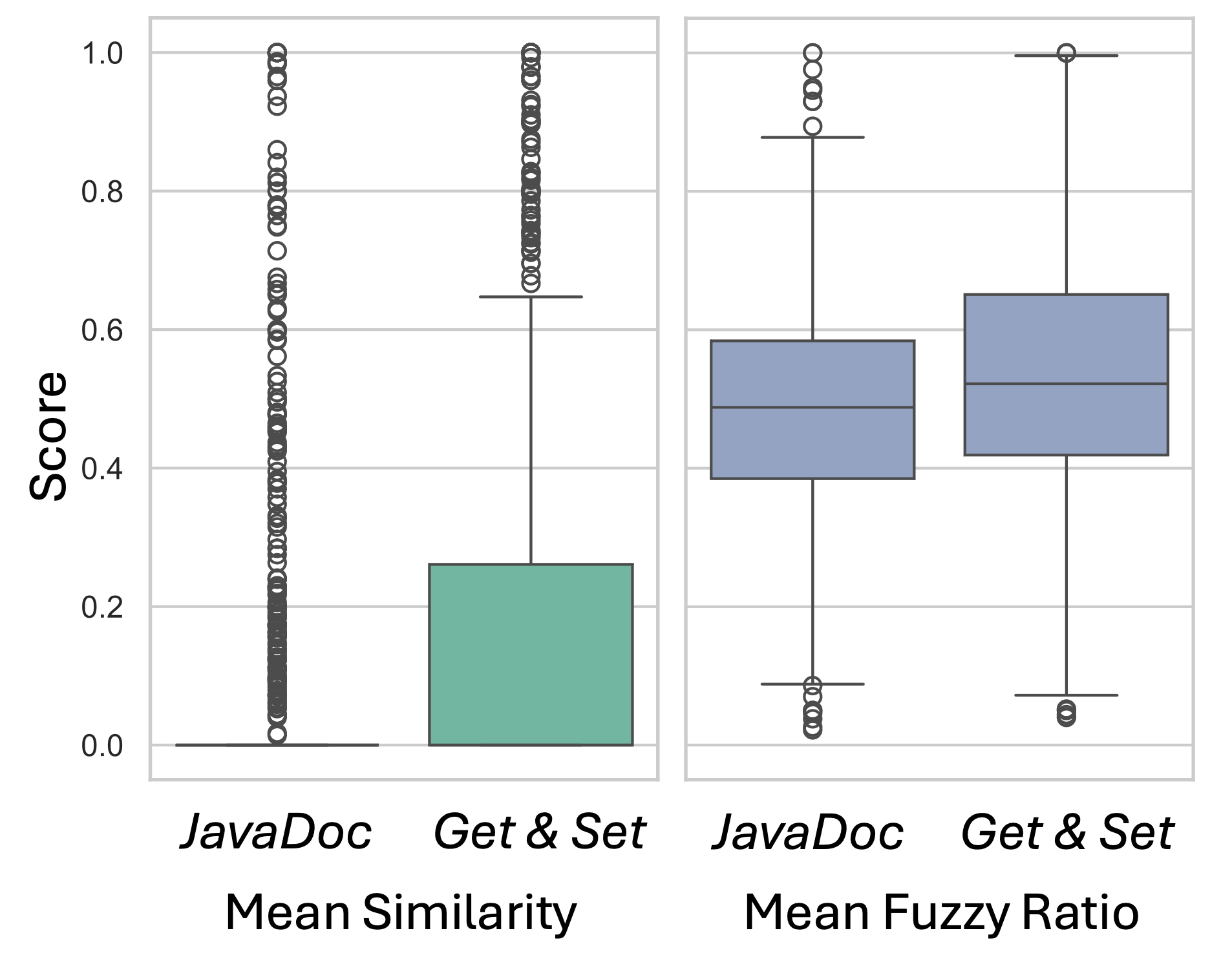} 
\vspace{-10pt} 

\caption{Similarity Scores for different contexts in the case of mean similarity.} \label{fig:similarity_javadoc_vs_access_in_mean_similarity} \end{figure}

\begin{figure}[h] \centering \includegraphics[width=0.4\textwidth]{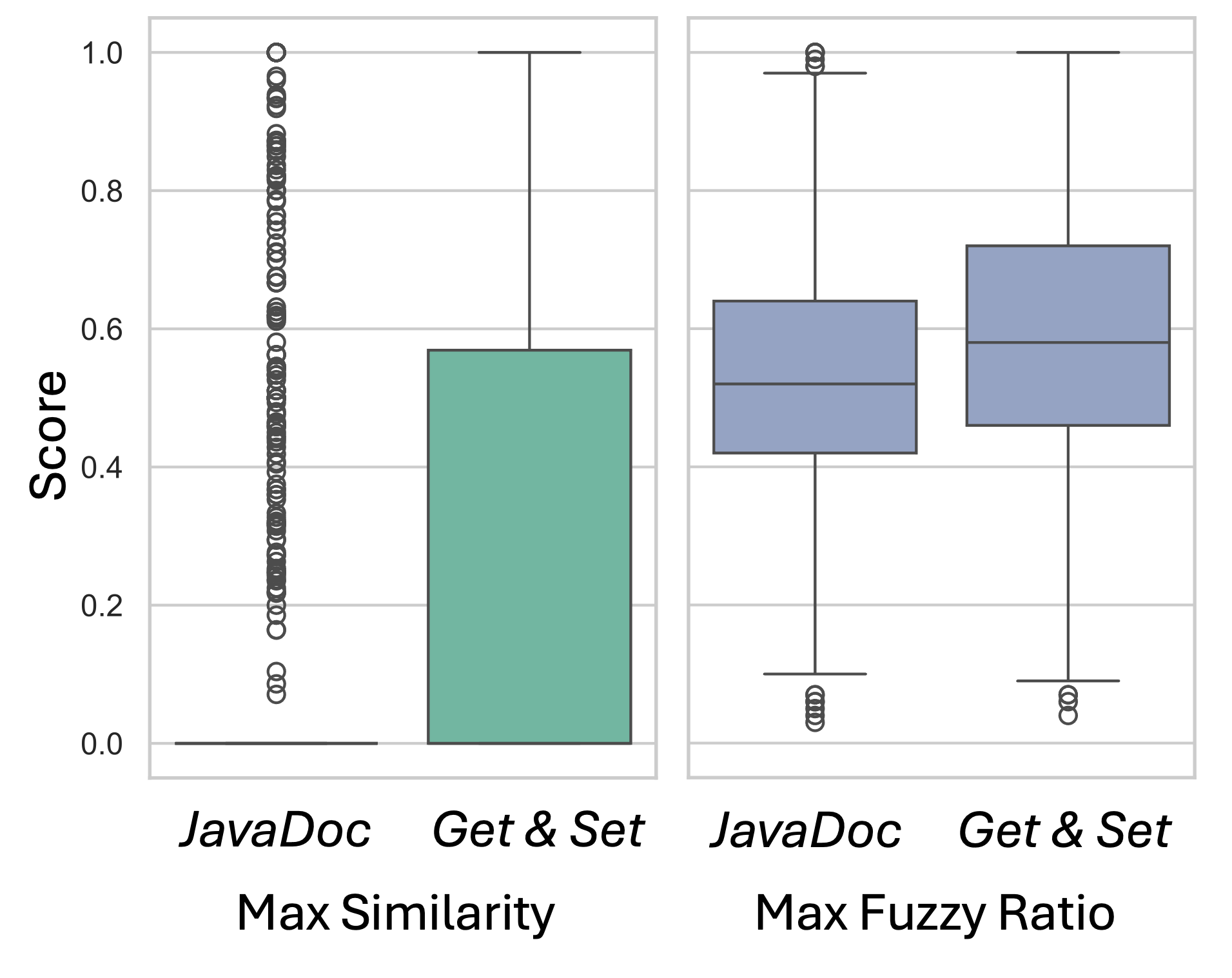} 
\vspace{-10pt} 

\caption{Similarity Scores for different contexts in the case of max similarity.} \label{fig:similarity_javadoc_vs_access_in_max_similarity} \end{figure}

The results of the Wilcoxon signed-rank test confirm the statistically significant difference in all similarity scores between the two contexts ($p < 0.05$), with access methods leading to higher similarity. 
Additionally, the effect size indicates a large difference in mean and max similarity, and a medium difference in mean and max fuzzy ratio, indicating that the inclusion of access methods is not only statistically significant but also practically relevant for the generated similarity scores.

\begin{table}[ht]\begin{center}
\caption{Count of methods for similarity values over threshold} 

\begin{tabular}{c|c|c}
\toprule
\textbf{Similarity}             & \textbf{Mean Similarity} & \textbf{Max Similarity} \\
\textbf{Values}                 & \textbf{Frequency}       & \textbf{Frequency}      \\ \midrule
1.00                            & 10 (1.58\%)              & 34 (5.39\%)             \\
{[}0.95, 1.00{]}                & 8 (1.27\%)               & 22 (3.49\%)             \\
{[}0.90, 0.95{]}                & 7 (1.11\%)               & 10 (1.58\%)             \\
{[}0.85, 0.90{]}                & 6 (0.95\%)               & 14 (2.22\%)             \\
{[}0.80, 0.85{]}                & 11 (1.74\%)              & 11 (1.74\%)             \\
{[}0.75, 0.80{]}                & 9 (1.43\%)               & 15 (2.38\%)             \\
{[}0.70, 0.75{]}                & 10 (1.58\%)              & 12 (1.90\%)             \\ \midrule
\textbf{Total {[}0.70, 1.00{]}} & \textbf{61 (9.66\%)}     & \textbf{118 (18.70\%)}  \\ \bottomrule
\end{tabular} 
%\vspace*{1mm}

\label{tab:count_methods_similarity_access} \end{center}
\end{table}

We further examined the number of methods with mean and max similarity exceeding the threshold of 0.70.
The results, as detailed in Table~\ref{tab:count_methods_similarity_access}, indicate a significant increase in the number of methods that exceed the plagiarism threshold when using access methods as context. Specifically, the proportion of methods classified as potentially plagiarized in the mean similarity case increased from 4.12\% to 9.66\% (2,3X increase), and the percentage increased from 9.19\% to 18.70\% (2X increase) for max similarity, presenting a $2X$ increase factor. 

%These results show how even a context carrying little information may induce the generation of code recommendations that mirror existing copyleft code.  

\noindent \fbox{\parbox{0.98\columnwidth}{\textbf{Answer to RQ3} Including access methods in the context affects the likelihood of generating plagiarized code by a factor of about $2X$. This result further confirms how incrementally accepting recommendations of plagiarized code can expose 
developers to an increasingly higher risk of accumulating unwanted code protected by copyleft licenses in their implementation.}}

%These findings highlight the importance of context in influencing the likelihood of generating code that may infringe on existing licenses. The substantial increase in the number of methods exceeding the similarity threshold when using access methods suggests that a more comprehensive context not only enhances the quality of the generated code but also raises the risk of unintentional plagiarism.

\subsection{RQ4: Does adjusting the temperature parameter alter the likelihood of obtaining code protected by copyleft licenses?}

RQ4 investigates the role of temperature in the likelihood of generating code that may break copyleft licenses. We thus investigated the impact of reducing the temperature from 1 to 0, computing the difference in the mean and max similarity and fuzzy ratio for all the methods in the dataset.

\begin{figure}
    \centering
    \includegraphics[width=0.4\textwidth]{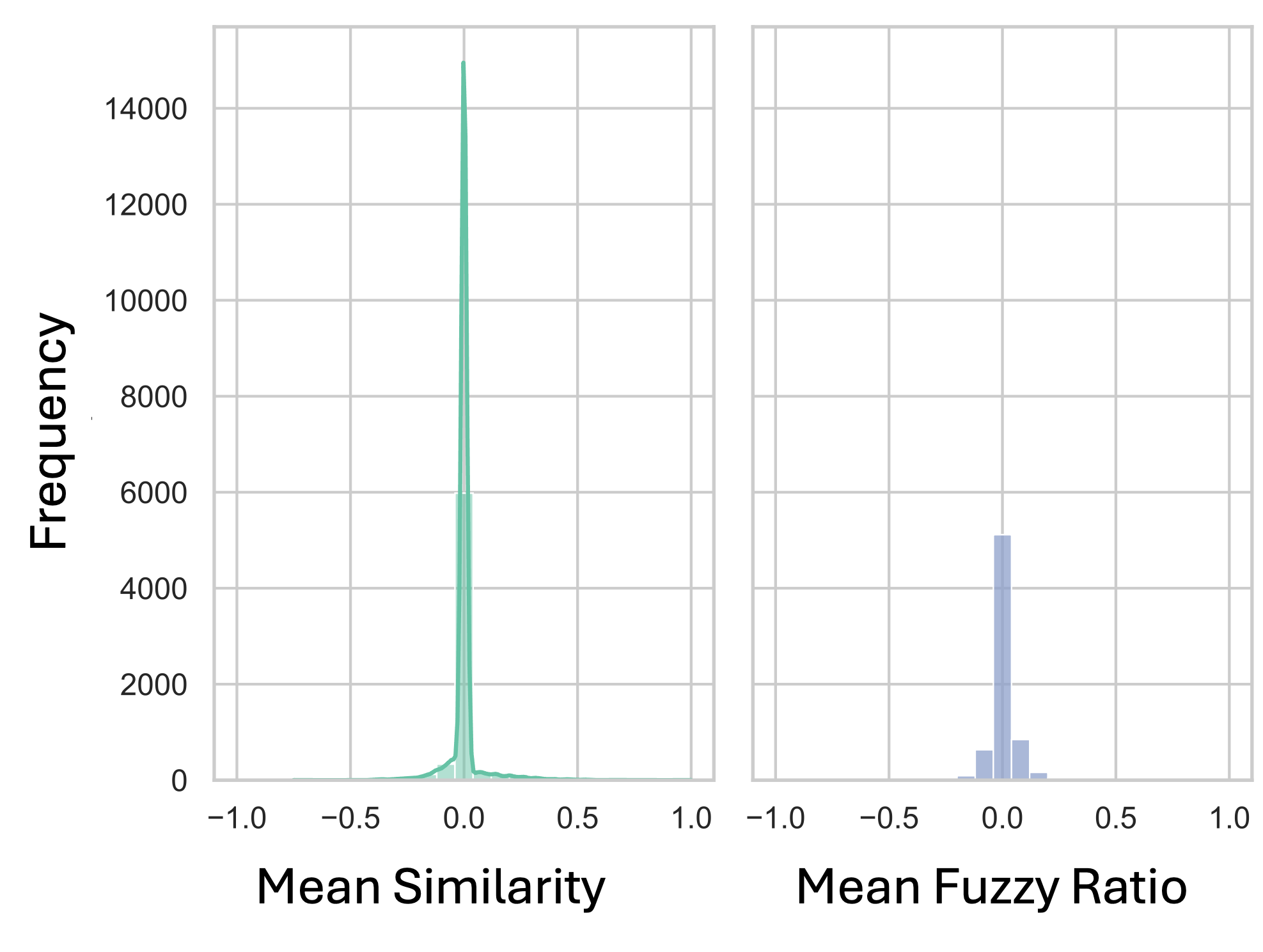}
    \vspace{-10pt} 
    
    \caption{Difference between temperature 0 and 1 in the case of mean similarity.}
    \label{fig:hist_diff_temperature_01_avg}
\end{figure}

\begin{figure}
    \centering
    \includegraphics[width=0.4\textwidth]{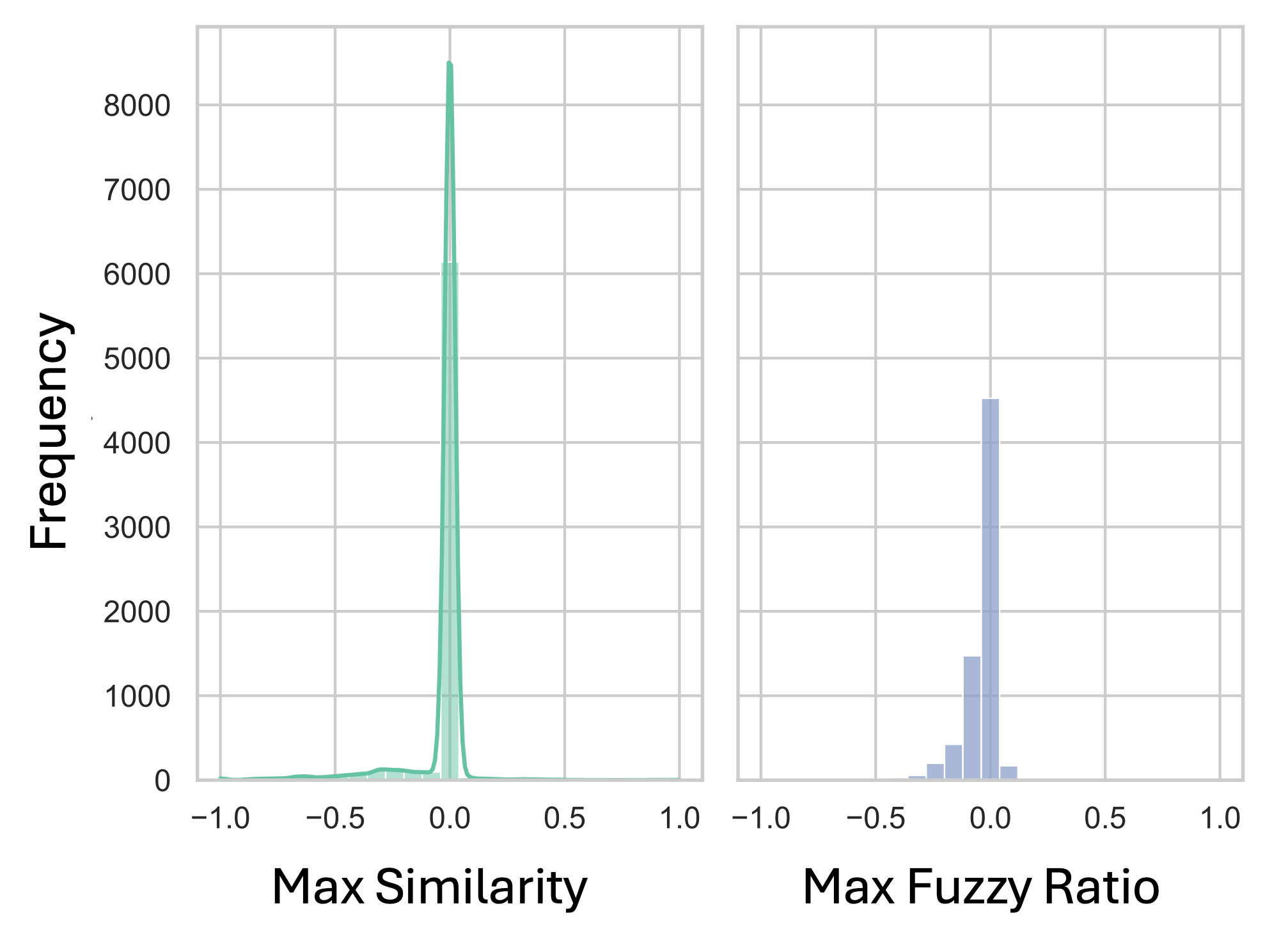}
    \vspace{-10pt} 
    
    \caption{Difference between temperature 0 and 1 in the case of max similarity.}
    \label{fig:hist_diff_temperature_01_max}
\end{figure}

The results of this comparison are visually represented in Figure \ref{fig:hist_diff_temperature_01_avg} for mean similarity and fuzzy ratio, and Figure \ref{fig:hist_diff_temperature_01_max} for max similarity and fuzzy ratio. We can observe that for most methods, the responses yielded similar similarity values (i.e., most of the differences close to 0). Specifically, 83.11\% of the methods reported the same value for mean similarity, while 86.11\% reported the same value for max similarity. For the remaining methods, the difference is very close to zero, with a slight bias toward greater plagiarism at temperature 1. However, the distribution of values is dense around 0 and balanced, suggesting that differences are not substantial. This is confirmed by a two-tailed Wilcoxon signed-rank test ($\alpha = 0.05$), indicating significance for max similarity ($\textit{p-value} = 8.66 \times 10^{-105}$), max fuzzy ratio ($\textit{p-value} = 0$) and mean fuzzy ratio ($\textit{p-value} = 3.91 \times 10^{-08}$), but not for mean similarity ($\textit{p-value} = 0.61$). 

The difference in the fuzzy ratio can be likely attributed to the sensitivity of the metric to small syntactic variations of the code. The difference in max similarity and not for mean similarity shows a mild impact of temperature affecting results in terms of the highest similarity that can be observed across five repetitions, but not affecting the mean value. 

The distributions that are slightly skewed towards positive values show that code with higher similarity is generated with a temperature equal to $1$. The percentage of potentially plagiarized methods is 1.48\% with temperature 1 and 1.89\% with temperature 0, according to mean similarity. The potentially plagiarized methods raise to 2.05\% for temperature equal to 0 and 3.35\% for temperature equal to 1, according to max similarity. Again, these values confirm the mild impact of temperature. %Despite the significant differences in similarity values, the number of methods with mean similarity above the threshold of 0.70 did not vary significantly, which can be attributed to greater variation in values below this threshold.

We finally investigated if the maximum temperature value can reduce the similarity of code compared to the default value of temperature equal to $1$. For this investigation, we used the methods that result in a similarity higher than 0.9 in at least one of the requests for a temperature equal to 1, focusing on the methods that are most exposed to the risk of plagiarism. 

\begin{figure}
    \centering
    \includegraphics[width=0.4\textwidth]{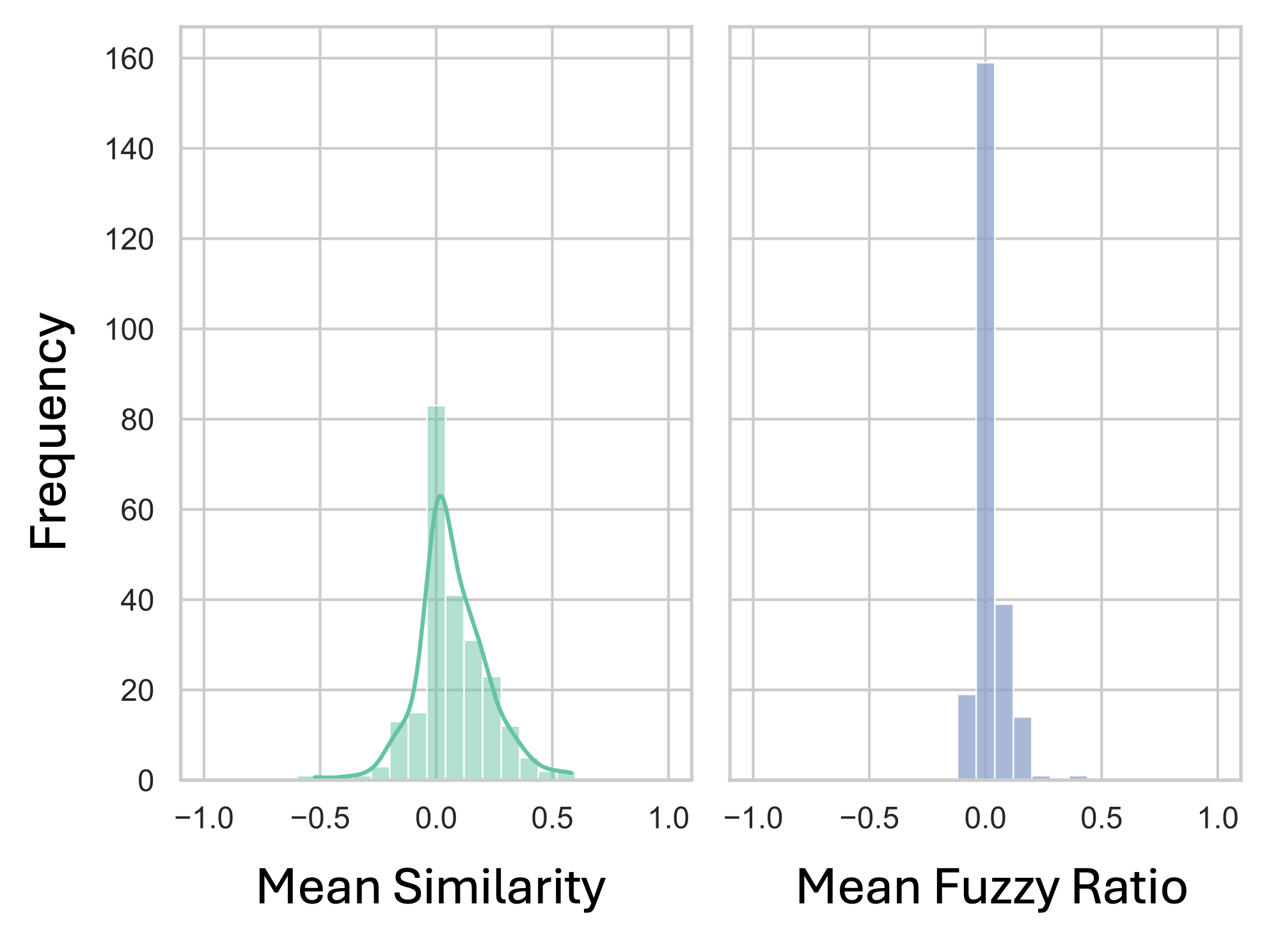}
    \vspace{-10pt} 
    
    \caption{Difference between temperature 1 and 2 in the case of mean similarity.}
    \label{fig:hist_diff_temperature_12_avg}
\end{figure}

\begin{figure}
    \centering
    \includegraphics[width=0.4\textwidth]{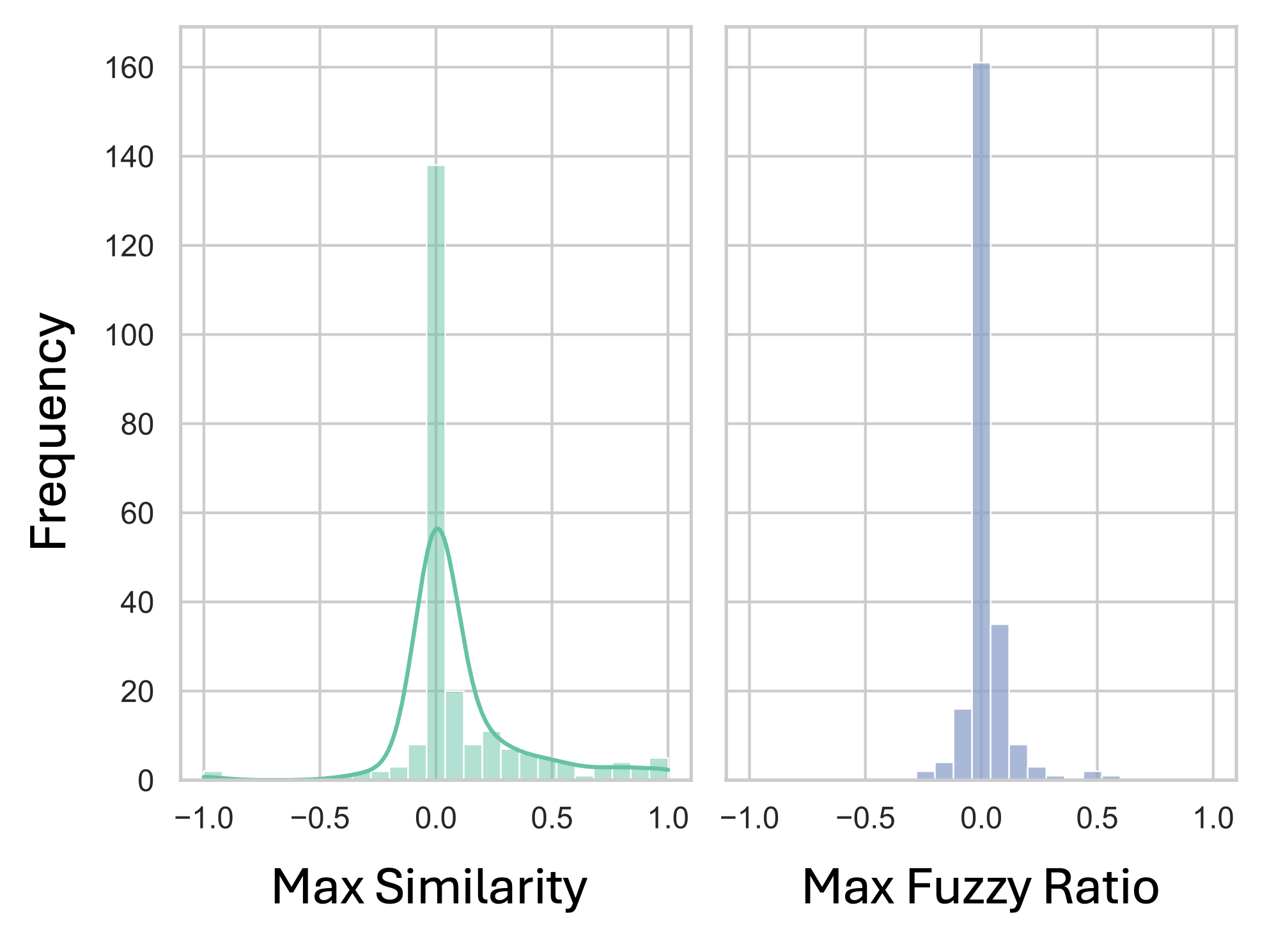}
    \vspace{-10pt} 
    
    \caption{Difference between temperature 1 and 2 in the case of max similarity.}
    \label{fig:hist_diff_temperature_12_max}
\end{figure}

The results of this comparison are illustrated in Figure~\ref{fig:hist_diff_temperature_12_avg} for mean similarity and Figure~\ref{fig:hist_diff_temperature_12_max} for max similarity.
In both cases, we observe a greater number of positive differences, indicating that responses obtained with temperature 2 are generally less similar to the original code than those with temperature 1. Specifically, a substantial percentage of methods reported lower similarity values with temperature 2, highlighting the effectiveness of the higher temperature in promoting variability and originality.

%In this analysis, we find that the responses generated at temperature 2 exhibit a significant increase in variability compared to those at temperature 1. This suggests that raising the temperature parameter encourages a broader range of outputs, potentially reducing the likelihood of similarity to existing content. 

The analysis using the two-tailed Wilcoxon signed-rank test ($\alpha = 0.05$) confirmed that mean and max similarity differences are significant (p-values equal to $9.94 \times 10^{-13}$ and $4.04 \times 10^{-10}$). Effect sizes of 0.4980 and 0.5804, further support this finding, capturing how responses generated at temperature 2 are largely and moderately different from those generated at temperature 1, for max and mean similarity respectively.   

%It is also important to note that when considering fuzzy ratio values, small effects are observed in both cases of similarity. This may suggest that, although the fuzzy ratio can detect syntactic variations, its sensitivity to such variations is insufficient to highlight significant differences in terms of originality. 

Regarding the number of methods above the threshold, using temperature equal to $2$, they decreased from 84 to 63 for mean similarity and from 156 to 132 for max similarity, with a reduction percentage of 9.01 and 10.31, respectively. 

%noted that for mean similarity, the percentage of potentially plagiarized methods decreased from 36.06\%  with temperature 1 to 27.04\% with temperature 2. Similarly, for max similarity, the percentage of responses identified as potentially plagiarized dropped from 66.95\% with temperature 1 to 56.65\% with temperature 2. This trend underscores the effectiveness of higher temperature settings in producing more original outputs, as evidenced by the significant reduction in the number of methods exceeding the similarity threshold of 0.70.

We can thus conclude that higher temperature values (i.e., temperature equal to 2) promote the generation of original code, with lower temperature values (i.e., temperature between 0 and 1) having a mild effect on code similarity. However, the recommendation of using high-temperature values to avoid unwanted plagiarism has to be carefully balanced with settings recommended for code generation, where low-temperature values tend to produce better results~\cite{arora2024optimizing,ouyang2023empirical,liu2024your}.  

\noindent \fbox{\parbox{0.98\columnwidth}{\textbf{Answer to RQ4} Temperature values of   $0$ and $1$ have a mild impact on the degree of similarity of the generated code, while high-temperature values, such as $2$, can promote the generation of new code less similar to known code.}}

\subsection{RQ5: Is ChatGPT aware of using protected code and can it avoid plagiarism when explicitly requested?}

RQ5 explores whether ChatGPT can avoid generating plagiarized code when explicitly asked to not reuse existing implementations in the prompt. We tested this prompt on the generation of the 239 methods that \textcolor{review}{exceeded a similarity threshold of 0.90, measured by the JPlag's max similarity metric.}
%exceeded a similarity threshold of 0.90 in RQ1. 

\begin{figure}[h]
    \centering
    \includegraphics[width=0.4\textwidth]{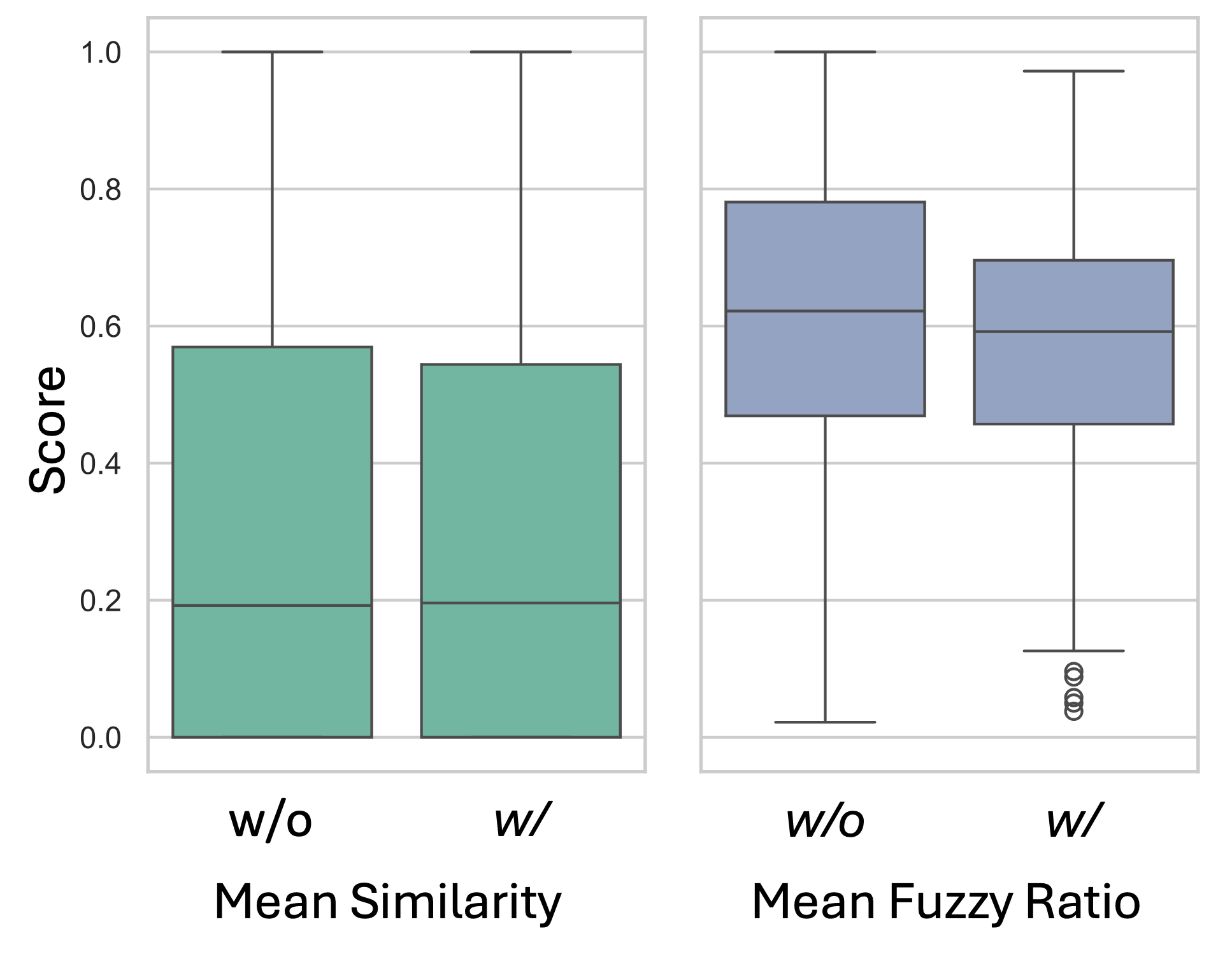}
    \vspace{-10pt} 
    
    \caption{Similarity scores generated without and with explicit instructions to avoid plagiarism in the case of mean similarity.}
    \label{fig:mean_similarity_with_instruction}
\end{figure}

\begin{figure}[h]
    \centering
    \includegraphics[width=0.4\textwidth]{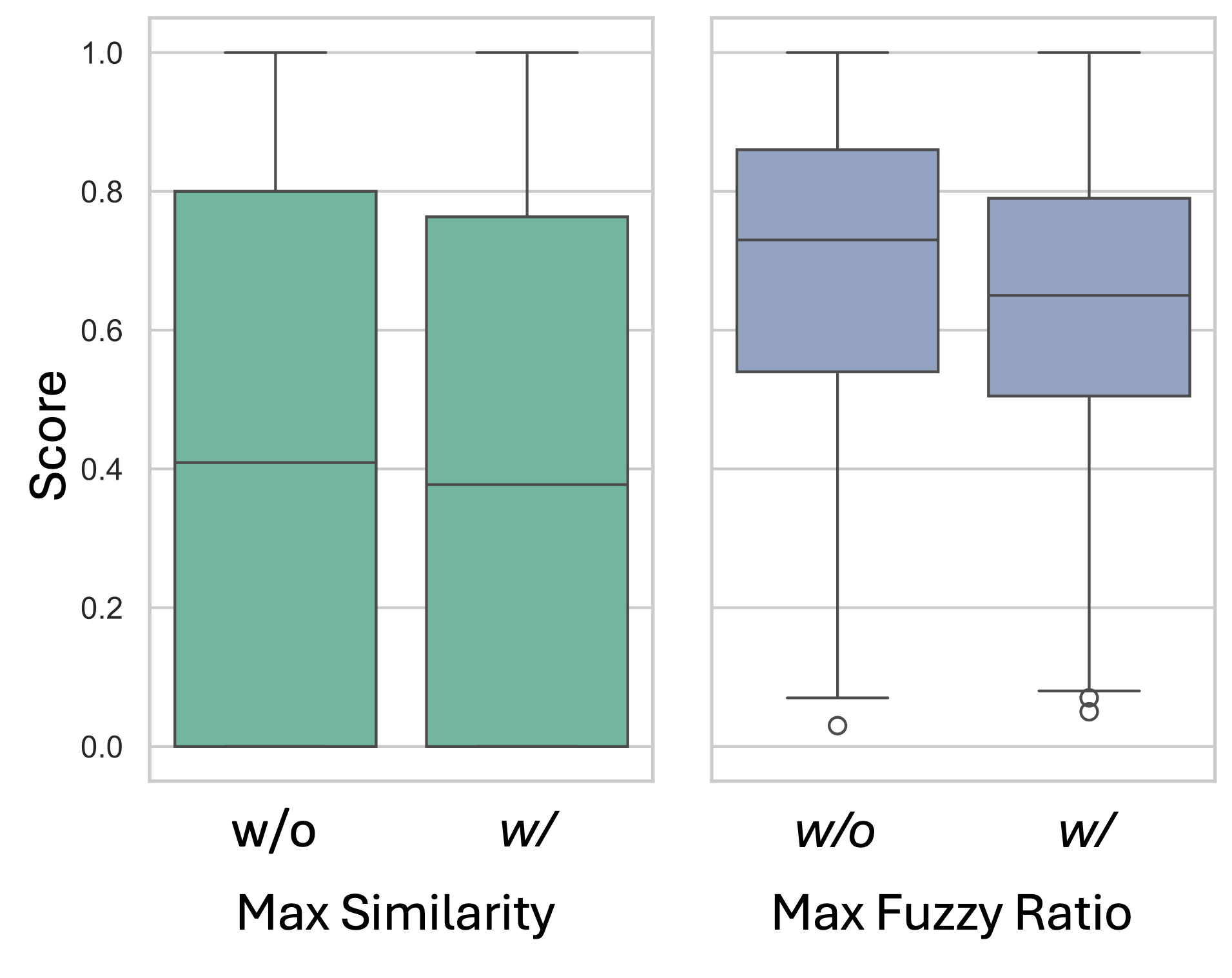}
    \vspace{-10pt} 
    
    \caption{Max similarity scores for outputs generated w/o explicit instructions.}
    \label{fig:max_similarity_without_instruction}
\end{figure}

Figures \ref{fig:mean_similarity_with_instruction} and \ref{fig:max_similarity_without_instruction} illustrate the similarity scores for outputs generated with and without the explicit request for originality. The data shows that the mean and max similarity scores remained relatively similar (scores close to 0), indicating that the explicit instruction did not lead to a substantial decrease in similarity. This is confirmed by a Wilcoxon signed-rank test ($\alpha=0.05$) that revealed no significant difference ($\textit{p-value} = 0.43$ for mean similarity and $\textit{p-value} = 0.98$ for max similarity) in similarity scores between the two conditions. 

This outcome indicates that the explicit request for unique code did not effectively alter the model's tendency to produce similar outputs to existing code.

%The results indicated that even with explicit instructions to avoid plagiarism, the similarity scores of the generated code did not show a significant reduction compared to outputs generated without such instructions. This finding suggests that while the model can produce original content, it may not fully comprehend the implications of plagiarism or the need for originality when prompted.

%The analysis of differences indicates that the difference distribution for both mean and max similarity cases closely aligns with a normal distribution, characterized by many methods exhibiting zero difference. Consequently, no significant differences are observed, regardless of whether a specific request for originality is made. 

%In summary, the experiments conducted for RQ5 demonstrate that the model's ability to distinguish between original and plagiarized content in the response remains limited. The similarity scores suggest that explicit instructions to avoid plagiarism do not significantly impact the originality of the generated code, raising concerns about the model's understanding of intellectual property issues.
\noindent \fbox{\parbox{0.98\columnwidth}{\textbf{Answer to RQ5} ChatGPT is not aware of reusing copyleft code and cannot be asked, through the prompt, to avoid reusing existing code in the responses.}}

\subsection{Threats to Validity}
The main threats to the internal validity of the results are about the selection of both the methods and the similarity metrics. 

Since we worked with a large pre-trained model, we cannot be sure about the cases that the model processed during training. We mitigated this threat by relying on publicly available information about the training date and the training process to ultimately select methods that are, almost certainly, processed by GPT-4-turbo. %Moreover, regardless of the possible memorization of samples, the results about the similarity of the recommendations with the copyleft code remain valid. 

Since the focus of the work is on the detection of code recommendations that expose developers to the risk of plagiarizing copyleft code, we relied on the use of a popular plagiarism detection tool, JPlag, to compute similarity. Although using alternative metrics the results may slightly change, we do not expect the results about nearly identical copies shall change, which are the cases discussed in this paper.

Since there exist additional sources of code than GitHub, our results should be intended as providing an under-approximation of the probability code protected by restrictive licenses is recommended by ChatGPT.

The main threats to the external validity are the generalization to other models and the language of our findings. 

In our study, we used GPT-4-turbo, which is the most advanced GPT model at the time of the study. Although we cannot claim our results to be valid for other LLMs, we do expect the observed degree of reuse of existing code to be an intrinsic characteristic of the architecture of the model, rather than a unique characteristic of the specific instance of the model used. Results may thus have validity going beyond the model studied in this paper. 

Finally, the study considers the case of Java code. Since our design does not include anything specific to Java, we do not expect the findings to be specific to Java. Of course, the likelihood of producing responses that resemble existing code may relate to the popularity of code samples available for the considered language, and languages that are more/less popular than Java may exhibit different patterns.

\section{Conclusions}
\label{sec:con}
Developers exploit AI assistants to speed up their work, interactively asking for code recommendations, such as the initial implementation of methods. However, AI assistants may inadvertently generate code that mirrors existing code protected by copyleft licenses, exposing developers who accept recommendations to the risk of reusing code without being aware of the legal implications of the reuse. 

In this paper, we systematically studied this phenomenon for ChatGPT and discovered that, although the risk of receiving such recommendations is mild %(3,35\% of the cases when ChatGPT is prompted with the method signature and Javadoc comment only) 
for individual requests, it might increase when a larger context is used. In particular, a matching set of access methods or a matching class code increases the chance of receiving the recommendation of copyleft code by a 2X and 5X factor, respectively. 
We also found that the temperature can be used to control this phenomenon, with high values reducing the likelihood of receiving recommendations with copyleft code. This recommendation has to be balanced with studies reporting that high-temperature values may decrease correctness. % of the recommendations. 

The results reported in this study open to several possible future works. For instance, the design of mechanisms to detect and control recommendations that expose developers to the risk of reusing copyleft code should be studied. Although it is hard to define a threshold that can define when reuse should be considered legally relevant, developers must be aware of the nature of the code they include in their software. Further, the study could be extended to other models, languages, and interaction modalities. %, to investigate if, how, and when LLMs may produce undesirable recommendations. 

\section*{Acknowledgment}
This work was partially supported by the MUR under the grant “Dipartimenti di Eccellenza 2023-2027" of the DISCo Department of the University of Milano - Bicocca; and by the Engineered MachinE Learning-intensive IoT systems (EMELIOT) national research project, funded by the MUR under the PRIN 2020 program (Contract 2020W3A5FY).

\bibliographystyle{IEEEtran}
  \bibliography{main}

%\vspace{12pt}

\end{document}